\documentclass[prb,twocolumn,footinbib,preprintnumbers,amsmath,amssymb,superscriptaddress,helvetica]{revtex4} 
\usepackage[T1]{fontenc}
\usepackage[utf8]{inputenc}
\usepackage[german,english]{babel}
\usepackage{epsfig}
\usepackage{amssymb}
\usepackage{verbatim}
\usepackage{amsmath}
\usepackage{braket}
\usepackage{bm}
\usepackage{mathtools}

\usepackage{xcolor}
\usepackage{times}
\usepackage[colorlinks,bookmarks=false,citecolor=blue,linkcolor=red,urlcolor=blue]{hyperref}

\usepackage{soul}

\newcommand{\beq}{\begin{equation}}
\newcommand{\eeq}{\end{equation}}
\newcommand{\bea}{\begin{eqnarray}}
\newcommand{\eea}{\end{eqnarray}}

\newcommand{\fsl}[1]{\ensuremath{\mathrlap{\!\not{\phantom{#1}}}#1}}

\begin{document}
\setstcolor{red}
\title{Lattice gauge theories and string dynamics in Rydberg atom quantum simulators}

\author{Federica M.~Surace}
\affiliation{SISSA -- International School for Advanced Studies, via Bonomea 265, 34136 Trieste, Italy.}
\affiliation{ICTP -- International Center for Theoretical Physics, Strada Costiera 11, 34151 Trieste, Italy.}
\author{Paolo P. Mazza}
\affiliation{SISSA -- International School for Advanced Studies, via Bonomea 265, 34136 Trieste, Italy.}
\affiliation{INFN, Sezione di Trieste, via Bonomea 265, 34136 Trieste, Italy.}
\author{Giuliano Giudici}
\affiliation{SISSA -- International School for Advanced Studies, via Bonomea 265, 34136 Trieste, Italy.}
\affiliation{ICTP -- International Center for Theoretical Physics, Strada Costiera 11, 34151 Trieste, Italy.}
\affiliation{INFN, Sezione di Trieste, via Bonomea 265, 34136 Trieste, Italy.}
\author{Alessio Lerose}
\affiliation{SISSA -- International School for Advanced Studies, via Bonomea 265, 34136 Trieste, Italy.}
\affiliation{INFN, Sezione di Trieste, via Bonomea 265, 34136 Trieste, Italy.}
\author{Andrea Gambassi}
\affiliation{SISSA -- International School for Advanced Studies, via Bonomea 265, 34136 Trieste, Italy.}
\affiliation{INFN, Sezione di Trieste, via Bonomea 265, 34136 Trieste, Italy.}
\author{Marcello Dalmonte}
\affiliation{SISSA -- International School for Advanced Studies, via Bonomea 265, 34136 Trieste, Italy.}
\affiliation{ICTP -- International Center for Theoretical Physics, Strada Costiera 11, 34151 Trieste, Italy.}


\begin{abstract}
Gauge theories are the cornerstone of our understanding of fundamental interactions among particles. Their properties are often probed in dynamical experiments, such as those performed at ion colliders and high-intensity laser facilities. Describing the evolution 
of these strongly coupled systems is a formidable challenge for classical computers, and represents one of the key open quests for quantum simulation approaches to particle physics phenomena.
In this work, we show how recent experiments done on Rydberg atom chains naturally realize the real-time dynamics of a lattice gauge theory at system sizes at the boundary of classical computational methods.
We prove that the constrained Hamiltonian dynamics induced by strong Rydberg interactions maps exactly onto the one of a $U(1)$ lattice gauge theory. 
Building on this correspondence, we show that the recently observed anomalously slow dynamics corresponds to a string-inversion mechanism, reminiscent of the string-breaking typically observed 
in gauge theories. This underlies the generality of this slow dynamics, which we illustrate in the context of one-dimensional quantum electrodynamics on the lattice.
Within the same platform, we propose a set of experiments that generically show long-lived oscillations, including the evolution of particle-antiparticle pairs, and discuss how a tuneable topological angle can be realized, further affecting the dynamics following a quench.
Our work shows that the state of the art for quantum simulation of lattice gauge theories is at 51 qubits, and connects the recently observed slow dynamics in atomic systems to archetypal phenomena in particle physics. 
\end{abstract}

\maketitle

\section{Introduction}
Lattice gauge theories (LGTs)~\cite{Wilson74} represent one of the most successful framework for describing 
fundamental interactions within the standard model of particle physics. Numerical simulations of their Euclidean formulation~\cite{Montvay1994} have shed light on paradigmatic equilibrium properties of 
strong interactions, including the low-lying spectrum of quantum chromodynamics~\cite{RevModPhys.84.449}, and the nature of its phase diagram~\cite{Fukushima_2010,deTar2015}. Non-
equilibrium properties, instead, are 
a notable challenge~\cite{calzetta_book}, 
due to the lack of generically applicable methods to simulate the real-time dynamics 
of extended, strongly interacting systems~\cite{Wiese:2013kk}.
This has stimulated an intense theoretical activity aimed at quantum simulating LGTs via
atomic quantum systems~\cite{Zohar2015,Dalmonte:2016jk,Preskill:2018aa}, leading to the first door-opener experimental realization in a system of four trapped ions~\cite{ExpPaper}. While such quantum simulators have already challenged the most advanced computational techniques in regard of condensed-matter motivated models~\cite{Trotzky2012,Bloch2012}, there is presently no experimental evidence that atomic systems can be used to simulate LGTs at large scales, nor that they can display physical phenomena with a direct counterpart in 
LGTs. 
This limitation stems from the very characteristic aspect that distinguishes LGTs from other statistical mechanics models, i.e, the presence of local constraints on the possible configurations, in the form of a Gauss law, 
which cannot be easily implemented in 
actual experimental realizations~\cite{Zohar2015,Dalmonte:2016jk}.

\begin{figure*}[t]
\begin{tabular}{cc}
\includegraphics[width=0.99\textwidth]{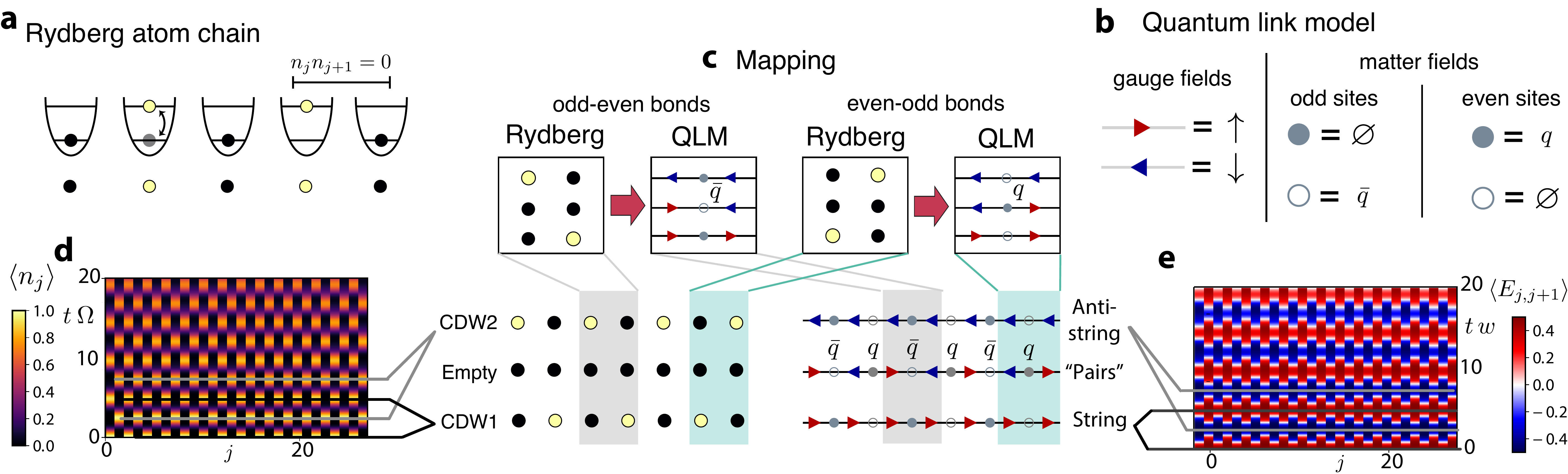}
\end{tabular}
\caption{%
Gauge-theory interpretation of Rydberg-atom quantum simulations.
{\bf a}: Schematics of a Rydberg atom chain. Each potential well of the optical lattice hosts a single atom, which can be either in the ground (black) or excited Rydberg (yellow) state. The two levels are coupled by a 
laser field. The Rydberg blockade prevents the simultaneous excitations of neighboring atoms. {\bf b}: Degrees of freedom of a $U(1)$ 
LGT in the spin-$1/2$ quantum link model (QLM) formulation. Gauge fields are represented by spin variables residing on links. Matter fields are represented by Kogut-Susskind fermions: an occupied site corresponds to the vacuum on odd sites, and to a quark $q$ on even sites. An empty site, instead, to the vacuum on even sites and to an anti-quark $\bar q$ on odd sites. {\bf c}: Mapping between Rydberg-blockaded states and configurations of the electric field constrained by the Gauss law in the QLM.
Due to the staggered electric charge, the allowed configurations of the electric field depend on the link, as illustrated. 
The two so-called charge-density wave configurations ``CDW1'' and ``CDW2'' of the Rydberg-atom arrays are mapped onto the ``string'' and ``anti-string'' states, respectively, characterized by uniform rightward or leftward electric fluxes. 
The empty 
configuration with all Rydberg atoms in their ground state is mapped to a state filled by adjacent particle-antiparticle pairs. 
{\bf d}: Time evolution of the Rydberg array governed by the effective Hamiltonian $H_{\text{FSS}}$ in Eq. \eqref{eq:fss}, starting from the CDW1 state. 
The plot shows the space and time resolved population $\langle n_j \rangle$ of the excited Rydberg atoms.
{\bf e}: Evolution of the expectation value of the electric field operator $\hat{E}_{j,j+1}$ in the QLM. These dynamics map exactly onto the ones shown in panel {\bf d} via the mapping illustrated in panel {\bf c}.
The thin lines highlight the oscillation between CDW1, CDW2 (left, 
bottom of panel {\bf c}) or string and anti-string (right) states.
In these simulations, $L=24$ and $\delta=m=0$.
}
\label{fig:cartoon}
\end{figure*}
%

Here, we show that (1+1)-dimensional LGTs akin to quantum electrodynamics are naturally realized in state-of-the-art experiments with Rydberg atom arrays~\cite{Bernien2017,Barredo:2018aa}. In particular, we show how the dynamics of Rydberg excitations in these chains is exactly mapped onto a spin-$1/2$ quantum link model (QLM), a $U(1)$  LGT where the gauge fields span a finite-dimensional Hilbert space, 
equivalent to a lattice Schwinger model in the presence of a topological term~\cite{Coleman1976239}.  
The key element of our mapping, which is schematically illustrated in Fig.~\ref{fig:cartoon}, is that gauge invariance has a natural counterpart in the Rydberg blockade mechanism, which constrains the Hilbert space in the same way as Gauss law does in gauge theories. 
This provides an immediate interpretation of the recent experiment with Rydberg-blockaded atom arrays in Ref.~\onlinecite{Bernien2017} as the first large-scale quantum simulation of a LGT 
at the edge of classical computational methods~\cite{Wiese:2013kk}.

From a theoretical viewpoint, the mapping offers a hitherto unexplored perspective on the anomalously slow relaxation recently observed in experiments: the long-lived oscillations in the population of excited Rydberg atoms correspond to a 
string inversion, a phenomenon which is directly tied to string breaking~\cite{calzetta_book,PhysRevD.71.114513,Hebenstreit2013} prototypical of gauge theories including dynamical matter (cf. Fig.~\ref{fig:cartoon}{\bf d} and \ref{fig:cartoon}{\bf e}). The mapping indicates that this phenomenon has a natural interpretation in the LGT framework, and suggests the occurence of slow dynamics in other $U(1)$ gauge theories, such as higher-spin 
QLMs~\cite{Chandrasekharan1997}, Higgs theories~\cite{Kuno_2015,PhysRevD.95.094507}, and the Schwinger model~\cite{Schwinger1,KogutSusskindFormulation}. These theories have been widely discussed in the context of Schwinger pair production taking place at high-intensity laser facilities, thus providing a highly unexpected, direct link between apparently unrelated experimental platforms~\cite{HebenstreitBerges2013, Hebenstreit2013, Buyens:2015lq, Pichler:2016it,Rajagopal1993}.

We discuss the generality of this type of quantum evolution by extending our analysis to other relevant instances of "slow dynamics", characterized by 
the absence of relaxation on all time scales corresponding to any microscopic coupling present in the system. 
As initial states, we focus on those consisting of 
particle-antiparticle pairs, corresponding to regular configurations of the Rydberg-atom arrays with 
localized defects, which are accessible within the setup of Ref.~\onlinecite{Bernien2017}. 
We show that these defects 
propagate ballistically with long-lived coherent interference patterns. This behavior is found to be governed by special bands of highly excited eigenstates characterized by a regularity in the energy-momentum dispersion relation. 
These findings open up a novel perspective which complements and extends towards gauge theories 
recent approaches to slow relaxation 
in Rydberg-blockaded atomic chains~\cite{Papic2018short, Papic2018long, Motrunich2018, Pichler2018TDVP, Choi2018SU2, Khemani2018}.

\vspace{0.2cm} 
\section{Rydberg atom arrays}
We are interested here in the dynamics of a one-dimensional array of $L$ optical traps, each of them hosting a single atom, as schematically illustrated in Fig.~\ref{fig:cartoon}{\bf a}. The atoms are trapped in their electronic ground state (black circle), denoted by $\Ket{\downarrow}_j$, where $j$ numbers the trap. These ground states are quasi-resonantly coupled to a single Rydberg state, i.e., a highly excited electronic level, denoted by $\Ket{\uparrow}_j$. The 
dynamics of this chain of qbits $\{\Ket{\uparrow,\downarrow}_j\}_{j=1,\ldots,L}$ 
is governed by the following Ising-type Hamiltonian~\cite{Bloch2012,Lesanovsky2012}:
 \beq
 \label{eq:Ryd}
 \hat H_{\text{Ryd}} = \sum_{j=1}^{L}(\Omega \,\hat\sigma^x_j +\delta\, \hat\sigma^z_j) + \sum_{j < \ell=1}^{L} V_{j,\ell} \hat n_j\hat n_\ell,
 \eeq
where $\hat\sigma^\alpha_j$ are Pauli matrices at 
site $j$, the operator $\hat n_j = (\hat \sigma^z_j+1)/2$ signals the presence of a Rydberg excitation at site $j$,  $2\Omega$ and $2\delta$ are the Rabi frequency and the detuning of the laser excitation scheme, respectively, and $V_{j,\ell}$ describes the interactions between atoms in their Rydberg states at sites $(j,\ell)$. For the cases of interest here, this interaction is strong at short distances and decays as $1/|j-\ell|^6$ at large distances. The dynamics described by $\hat H_{\text{Ryd}}$ has already been realized in several experiments utilizing either optical lattices or optical tweezers~\cite{Barredo:2018aa,Zeiher:2017aa,Bernien2017}. In particular, Ref.~\onlinecite{Bernien2017}
investigated the case in which 
$V_{j,j+1}$ is much larger than all other energy scales 
of the 
system, resulting in the so-called Rybderg {\it blockade effect}: atoms on neighboring sites cannot be simultaneously excited to the Rydberg state, hence the constraint $\hat n_j\hat n_{j+1} = 0$. 

In this regime, the resulting Hamiltonian --- introduced by Fendley, Sengupta and Sachdev (FSS) in Ref.~\onlinecite{Fendley2004} --- is 
 \beq
 \label{eq:fss}
 \hat H_{\text{FSS}} = \sum_{j=1}^L \left(\Omega \, \hat \sigma_j^x + 2\delta\, \hat n_j\right),
 \eeq
where we neglect longer-range terms which do not affect qualitatively the dynamics. $\hat H_{\text{FSS}}$ 
acts on the constrained Hilbert space without double occupancies on nearest-neighbor sites, as illustrated in Fig.~\ref{fig:cartoon}{\bf a}.
As we show below, the direct connection between Rydberg atomic systems and gauge theories is indeed provided by this constraint at the level of the Hilbert space.

\vspace{0.2cm} 

\section{Rydberg blockade as a gauge symmetry constraint}

\label{sec:mapping}
We establish here the exact mapping between the FSS Hamiltonian in Eq.~\eqref{eq:fss} governing the dynamics of the Rydberg atom quantum simulator in Ref.~\onlinecite{Bernien2017} and a $U(1)$ LGT. 
The latter describes the interaction between fermionic particles, denoted by $\hat{\Phi}_j$ and residing on the  lattice site $j$, mediated by a $U(1)$ gauge field, i.e., the electric field $\hat E_{j,j+1}$, defined on lattice bonds, as depicted in Fig.~\ref{fig:cartoon}{\bf b}. We use here Kogut-Susskind (staggered) fermions~\cite{KogutSusskindFormulation}, with the conventions that holes on odd sites represent antiquarks $\bar q$, while particles on even sites represent quarks $q$. Their 
dynamics is described by:
\begin{multline}
   \hat{H}=-w\sum_{j=1}^{L-1}(\hat{\Phi}^\dag_j \hat U^{\mathstrut}_{j,j+1}\hat{\Phi}^{\mathstrut}_{j+1} + \textnormal{h.c.})+ m\sum_{j=1}^L(-1)^j\hat{\Phi}_j^\dag\hat{\Phi}^{\mathstrut}_j \label{Schwinger lattice}  \\
   + J\sum_{j=1}^{L-1} \hat E^{\mathstrut 2 }_{j,j+1},
\end{multline}
where the first term provides the minimal coupling between gauge and matter fields through the parallel transporter $\hat U_{j,j+1}$ with $[\hat E_{j,j+1},\hat U_{j,j+1}]=\hat U_{j,j+1}$, the second term is the fermion mass, and the last one is the electric field energy. 
The generators of the $U(1)$ gauge symmetry are defined as 
\begin{equation}
\hat G_j= \hat E_{j,j+1}-\hat E_{j-1,j}-\hat{\Phi}_j^\dagger \hat{\Phi}^{\mathstrut}_j+\frac{1-(-1)^j}{2},
\label{eq:Gauss}
\end{equation}
and satisfy $[\hat{H}, \hat G_j] = 0$, so that gauge invariant states $\Ket{\Psi}$ satisfy Gauss law 
$\hat G_j\Ket{\Psi}=0$ for all values of $j$. 
Restricting the dynamics to their subspace is by far the most challenging task for quantum simulators.

Different formulations of $U(1)$ LGTs 
are obtained for different representations of gauge degrees of freedom $\hat E_{j,j+1}$.
While in the standard Wilsonian formulation --- i.e., the lattice Schwinger model --- they 
span infinite-dimensional Hilbert spaces, here we first focus on the $U(1)$ QLM 
formulation~\cite{Chandrasekharan1997,QLink1}, where
they are 
represented by spin variables, i.e., $\hat E_{j, j+1}=\hat S^z_{j, j+1}$ and  $\hat U_{j, j+1}=\hat S^+_{j,j+1}$, 
so that $[\hat E_{j,j+1},\hat S^+_{j, j+1}] = \hat S^+_{j, j+1}$. As noted in Ref.~\onlinecite{Banerjee2012}, this formulation is particularly suited for quantum simulation purposes.

In the following, we 
consider the QLM with spin $S=1/2$, in which all the possible configurations of the electric field have the same electrostatic energy, rendering the value of $J$ inconsequential;  we anticipate that this model is equivalent to the lattice Schwinger model in the presence of a $\theta$-angle with $\theta=\pi$~\footnote{The similarity between the phenomenology of the two models was pointed out in Ref.~\onlinecite{Banerjee2012}. Here, we are instead interested in establishing an exact relation.}.
The Hilbert space structure following Gauss law is particularly simple in this case~\cite{Banerjee2012}: as depicted in Fig.~\ref{fig:cartoon}{\bf c}, for each block along the chain consisting of two 
electric fields neighbouring a matter field at site $j$,
there are only three possible states, depending on the parity of $j$. In fact, in a general (1+1)-dimensional $U(1)$ LGT, 
the configuration of the electric field 
along the chain determines the configuration of the charges via the Gauss law. 
Accordingly, $\hat H$ in Eq.~\eqref{Schwinger lattice}
can be recast into a form in which the matter fields $\hat \Phi_{j}$ are integrated out. 

We now provide a transformation which maps exactly the latter form into the FSS Hamiltonian \eqref{eq:fss}. 
The correspondence between the two 
Hilbert spaces is realized by identifying, alternately on odd and even lattice sites, the computational basis configurations of the atomic qubits allowed by the Rydberg blockade  with the classical configurations 
of the electric field allowed by the Gauss law (see Fig.~\ref{fig:cartoon}{\bf c}). 
In terms of the two Hamiltonians \eqref{eq:fss} and 
\eqref{Schwinger lattice}, this unitary transformation consists in identifying the operators $\hat \sigma_j^x \leftrightarrow 2\hat S_{j-1,j}^x$, $\hat \sigma_j^{y,z} \leftrightarrow (-1)^j2\hat S_{j-1,j}^{y,z}$ and the parameters $\Omega=-w$, $\delta=-m$. 
This mapping overcomes the most challenging task in quantum simulating gauge theories, by restricting the dynamics directly within the gauge-invariant Hilbert space. 
The only states that would violate Gauss law 
are nearest-neighbor occupied sites which are strongly suppressed by the Rydberg blockade and can
be systematically excluded via post-selection of the configurations. Beyond providing a direct link between Gauss law and the Rydberg blockade mechanism, the most important feature of the mapping is that, 
differently from other remarkable relations between $\hat{H}_{\text{FSS}}$ and lattice models with gauge symmetries~\cite{McCoy1983,Chepiga:aa},
it provides an immediate connections between Rydberg experiments and particle physics phenomena, as we describe below.

\begin{figure}[t]
\includegraphics[scale=0.45]{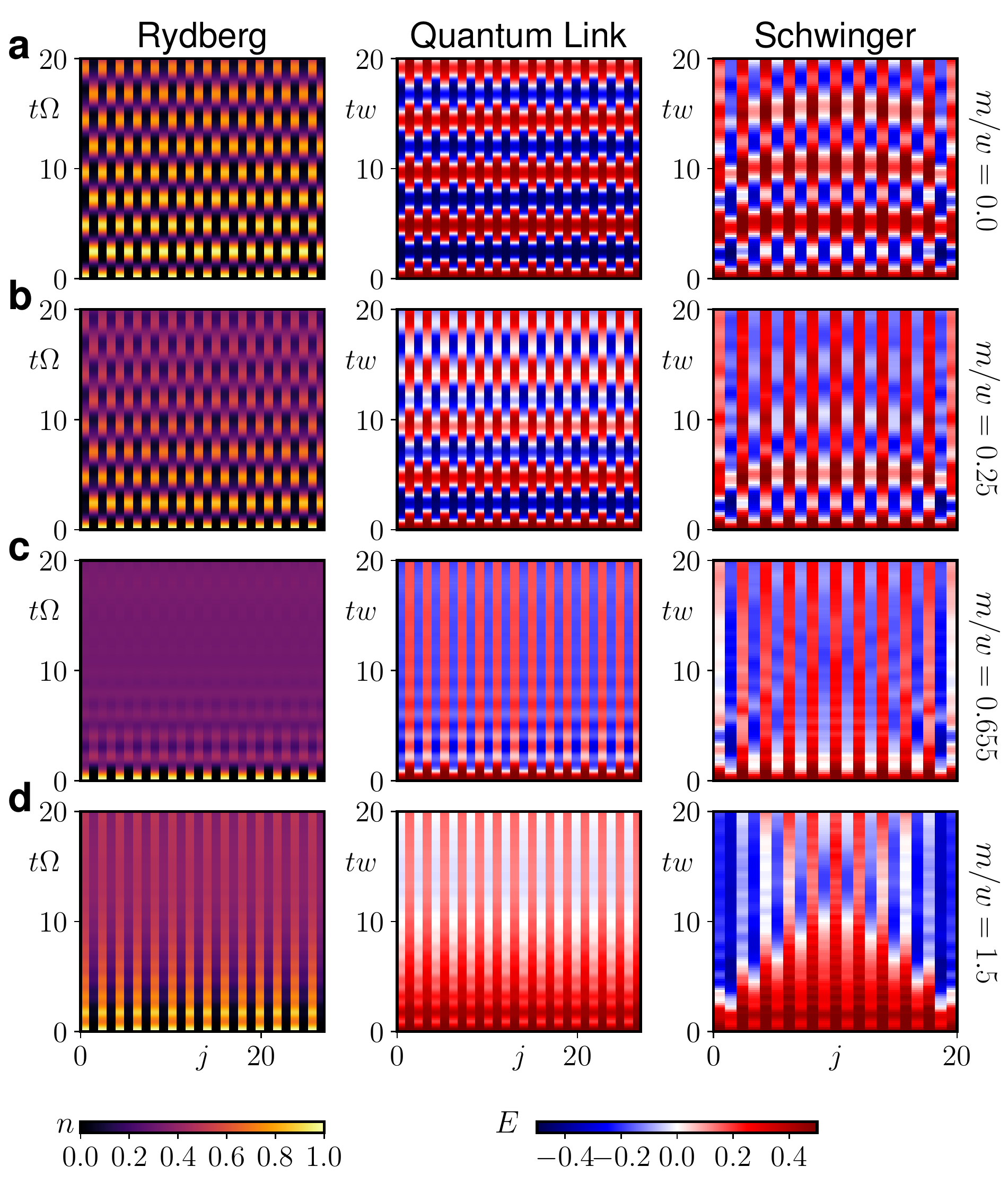}
\caption{%
Slow dynamics in Rydberg atoms, $U(1)$ quantum link model (QLM), and the lattice Schwinger model.
Coherent quantum evolution of (first column) the local Rydberg excitation density profile 
$n_j(t)=\langle \hat n_j(t) \rangle$ 
in the FSS model [see Eq.~\eqref{eq:fss}], starting from a charge-density wave, of the local electric field profile (second column) $E_{j,j+1}(t)=\langle \hat S^z_{j,j+1} (t) \rangle$ in the QLM, and (third column) $\langle \hat L_{j,j+1} (t)-\theta/(2\pi) \rangle$ (see further below in the main text) in the lattice Schwinger model [see Eq.~\eqref{Schwinger lattice}] with $J/w=1.5$ and $\theta=\pi$.
The four rows correspond to increasing values of the detuning $\delta$ (Rydberg) or, equivalently, of the particles mass $m=-\delta$ (QLM and Schwinger model).
Figures \ref{fig:cartoon}{\bf d} and \ref{fig:cartoon}{\bf e} correspond to the first two plots in panel {\bf a} here.
Data in the first and second columns are connected by a unitary transformation, while a remarkable similarity is manifest between the second and third column despite 
the larger Hilbert space of the gauge degrees of freedom in the Schwinger model.
The persistent string inversions observed within the symmetric phase with $m<m_c=0.655 |w|$ (first two lines) are suppressed as the quantum critical point is approached.
The dynamics in 
the third column feature edge effects due to the imposed open boundary conditions.
}
\label{fig:fss}
\end{figure}

\section{Real-time dynamics of lattice gauge theories in Rydberg atom experiments}

\subsection{Gauge-theory interpretation of slow dynamics}
The exact description of Rydberg-blockaded chains in terms of a $U(1)$ LGT allows us to shed a new light on the slow dynamics reported in Ref.~\onlinecite{Bernien2017},
by interpreting them in terms of well-studied phenomena in high-energy physics, related to the production of particle-antiparticle pairs after a quench akin to the Schwinger mechanism.

In the experiment, the system was initialized in a charge density wave state (CDW1 in Fig.~\ref{fig:cartoon}{\bf c}), and subsequently, the Hamiltonian 
was quenched, inducing slowly-decaying oscillations between CDW1 and CDW2. 
As shown in Fig.~\ref{fig:cartoon}{\bf c}, CDW1 and CDW2 are mapped onto the two states of the $S=1/2$-QLM with uniform electric field 
$\hat S^z_{j,j+1} = \pm 1/2$. 
The experimental results in Ref.~\onlinecite{Bernien2017} may thus be interpreted as the evolution starting from one of the two degenerate bare particle vacua $\Ket{0_{\pm}}$ (i.e, the vacua in the absence of quantum fluctuations, $w=0$)  of the gauge theory. 
In Fig.~\ref{fig:cartoon}{\bf d} and in the first column of Fig.~\ref{fig:fss}, we illustrate these dynamics as it would be observed in the excitation density $\langle n_j\rangle$ along the Rydberg-atom quantum simulators ("Rydberg") and compare it with that of the electric field $\langle E _{j,j+1}\rangle$ within its gauge-theory description ("QLM") in Fig.~\ref{fig:cartoon}{\bf e} and in the second column of Fig.~\ref{fig:fss}, respectively, utilizing exact diagonalization.

\begin{figure}[t]
\includegraphics[scale=0.44]{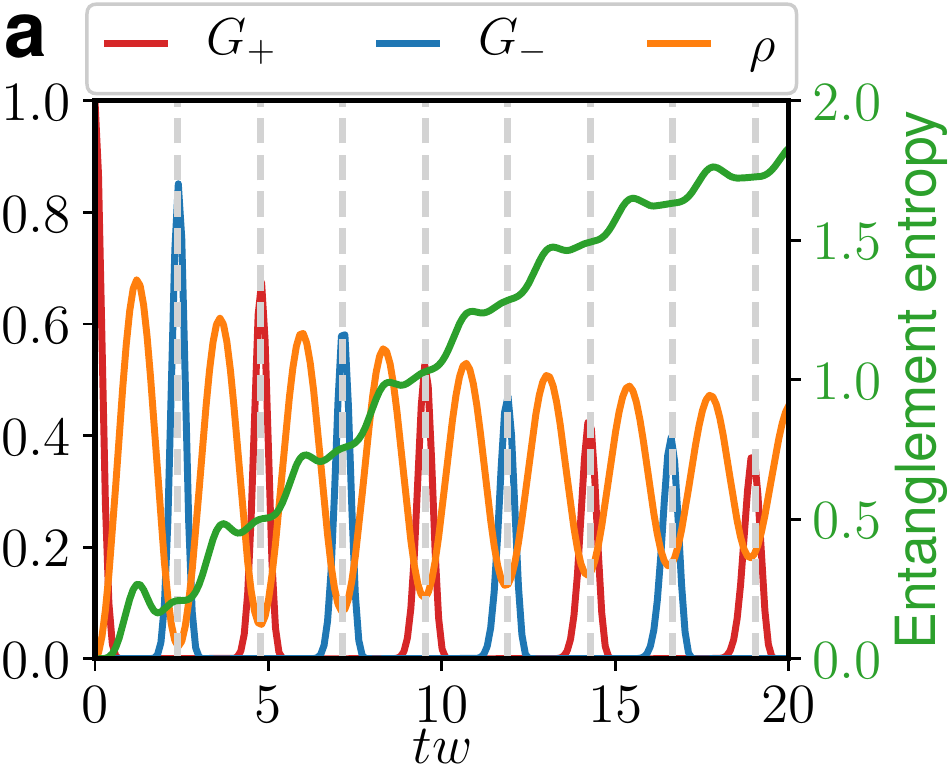} 
\includegraphics[scale=0.44]{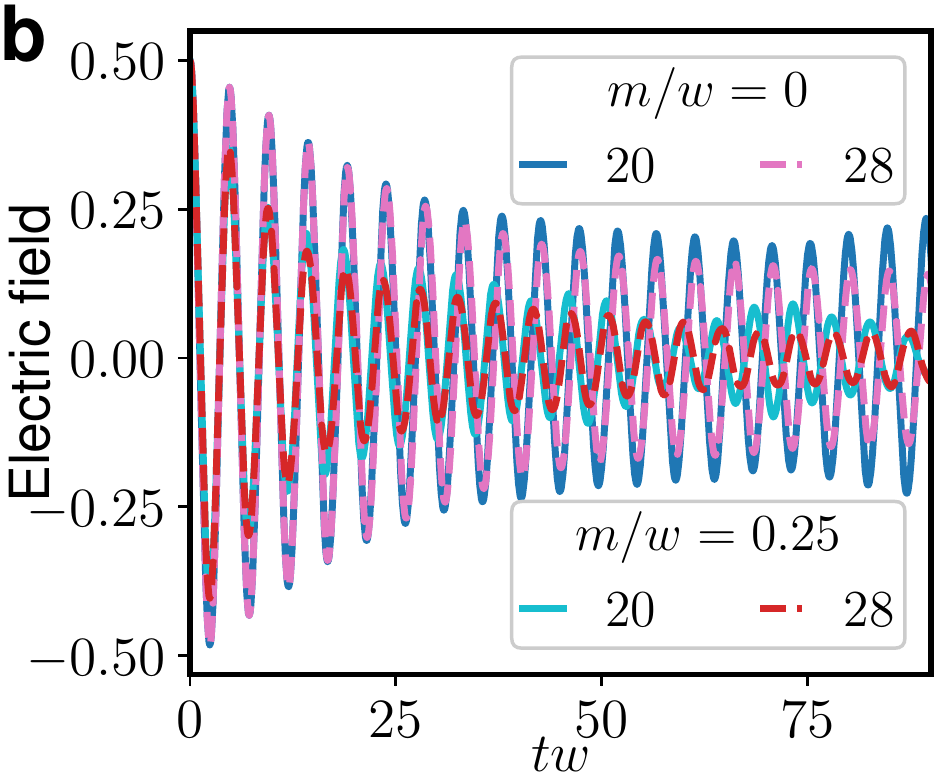} 
\caption{%
Characterization of slow dynamics in the FSS model.
{\bf a:} Hilbert space characterization of the persistent string inversions ($m=0$, $L=28$): alternating strong revivals of the overlaps $G_{\pm}(t)=|\braket{0_{\pm}|e^{-i\hat Ht}|0_+}|^2$ with the two bare vacuum states $\ket{0_{\pm}}$, corresponding to the two charge-density wave configurations of Rydberg-atom arrays. Both the total density $\rho = \langle \hat \rho_j\rangle$ of particle-antiparticle pairs, with 
$\hat\rho_j = (-1)^j \hat \Phi^\dagger_{j}\hat \Phi^{}_{j} +[1-(-1)^j]/2$ and the half-chain entanglement entropy (see the supplementary information) have regularly-spaced maxima between the peaks. {\bf b:} Persistent oscillations of electric field for two values of the mass and of the system size.
}
\label{fig:fsslongtime}
\end{figure}

The qualitative features of this evolution are strongly affected by quantum fluctuations, whose impact is quantified by the ratio between the coupling constant $w$ and the particles mass $m$. For small values of $m/w$ (first two lines in Fig.~\ref{fig:fss}), production of particle-antiparticle pairs occurs at a finite rate. We remark that this effect is reminiscent of the Schwinger mechanism~\cite{calzetta_book}, which however concerns pair creation from the true (and not the bare) vacuum. 
These particles get accelerated by the electric field and progressively screen it, until coherent pair annihilation takes place and eventually 
brings the system to a state with opposite electric flux.
This process, referred to as \textit{string inversion}, occurs 
several times in a coherent fashion, causing 
a dramatic slowdown of thermalization and of quantum information scrambling. As a further evidence, 
we compute both the total electric flux and the vacuum persistence amplitude 
(or Loschmidt echo), defined as $G_+(t)=|\braket{0_{+}|e^{-i\hat Ht}|0_+}|^2$, 
whose large value $\simeq 1$ 
was already noted in Ref.~\onlinecite{BanerjeeHeyl2018}. The anomalous long-lived oscillations of these quantities experimentally detected with Rydberg atom arrays in Ref.~\onlinecite{Bernien2017} show a clear analogy with several previous numerical studies of the real-time dynamics of higher-spin 
QLMs~\cite{Pichler:2016it} as well as of the Schwinger model~\cite{HebenstreitBerges2013,Kasper2015,Buyens:2015lq} and Higgs theories~\cite{PhysRevD.95.094507}. In addition, as noted in Ref.~\onlinecite{Banerjee2012}, the dynamics discussed here describes the coherent oscillations of the parity-symmetric  order parameter (in our case, $\langle \hat E_{j,j+1}\rangle$) as a function of time, reminiscent of the decay of a chiral condensate in QCD~\cite{Rajagopal1993}. We thus provide here a 
bridge among all these observations. 

However, if fermionic particles are sufficiently heavy, with $m/w$ exceeding a critical threshold, pair production is a virtual process and string inversion cannot be triggered, as shown in the third and fourth line of Fig.~\ref{fig:fss}. We find that this behavior is related to the quantum phase transition occurring in the FSS model 
at $\delta_c=-0.655|\Omega|$~\cite{Fendley2004}. This transition corresponds to the spontaneous breaking of the chiral symmetry in the LGT 
\eqref{eq:Gauss} at $m_c=0.655 |w|$~\cite{Rico:2014ek}. The four rows in Fig.~\ref{fig:fss} show the temporal evolution of the same initial uniform flux configuration (CDW or ``string'' in Fig.~\ref{fig:cartoon}{\bf c}) upon increasing values of the mass $m/w=0,0.25,0.655,1.5$ corresponding to the dynamics  ({\bf a}, {\bf b}) at $m<m_c$, 
({\bf c})  at the quantum critical point $m=m_c$, and ({\bf d}) at $m>m_c$.

Figure~\ref{fig:fsslongtime} further illustrates 
the appearance of string inversions for $m<m_c$ and the corresponding slow dynamics. Panel {\bf a} shows 
the long-lived revivals of the many-body wavefunction in terms of the evolution 
of the probability $G_{\pm}(t)$ of finding the system at time $t$ in the initial bare vacuum state $\ket{0_+}$ or in the opposite one $\ket{0_-}$, corresponding to $G_+$ or $G_-$, respectively, as well as in terms of the time-dependent density $\rho$ of particle-antiparticle pairs. The entanglement entropy of half system also displays an oscillatory behavior (see supplementary information).
Panel {\bf b} shows the scaling 
of the collective oscillations of the electric field with respect to the system size $L$, as well as their persistence with a small but non-vanishing fermion mass $m<m_c$.

\subsection{Slow dynamics in the 
Schwinger model}

\label{subs:latticeSchwinger}
The above phenomenology is not restricted to QLMs, 
but is expected to be a generic feature of LGTs including dynamical matter. 
We show this in the context of a Wilsonian
LGT, i.e., the lattice version of the Schwinger model in Eq.~\eqref{Schwinger lattice}. In this case, $\hat U_{j,j+1}=e^{i\hat \vartheta_{j,j+1}}$ are $U(1)$ parallel transporters with vector potential $\hat \vartheta_{j,j+1}$, the corresponding electric field operator is $\hat E_{j, j+1}=\hat L_{j, j+1}-\theta/(2\pi)$, where $\hat L_{j, j+1}$ have integer spectrum and $\theta/(2\pi)$ represents a uniform classical background field parameterized by the $\theta$-angle. Canonical commutation relations for the gauge degrees of freedom read $[\hat \vartheta_{j,j+1}, \hat L_{p, p+1}]= i\delta_{jp}$. In our numerical simulations, we utilize the spin formulation of the model obtained upon integration of the gauge fields under open boundary conditions ~\cite{Schwinger2,Banuls}.

We consider the case of a $\theta$-angle with $\theta=\pi$, such that two uniform field configurations have equal electrostatic energy. In the limit $J/w\rightarrow\infty$, the lattice Schwinger model is equivalent to the spin-$1/2$ QLM discussed above. We find evidence that the corresponding behaviour persists qualitatively down to $J\simeq w$, when the electrostatic energy competes with the matter-field interaction, as shown in the third column of Fig.~\ref{fig:fss}. Despite the strong quantum fluctuations allowed in principle by the exploration of a locally infinite-dimensional Hilbert space, a qualitative similarity --- which becomes a quantitative correspondence concerning the relationship between the coupling and the period of oscillation ---  with the case of the locally finite-dimensional Hilbert space of the QLM is manifest in the second column of Fig.~\ref{fig:fss}, related to the observed dynamics in Ref.~\onlinecite{Bernien2017}.

The generality of the occurrence of oscillations  which do not decay on time scales immediately related to the microscopic couplings points to a rather robust underlying mechanism. 
In fact, we suggest here that this behavior may arise from a universal field-theoretical description of the nonequilibrium dynamics of states possessing a well-defined continuum limit.
Concerning the $U(1)$ LGTs discussed in this work, the reference continuum field-theory description is provided by the Schwinger model, representing quantum electrodynamics in one spatial dimension \cite{}. In the massless limit $m=0$, this model can be exactly mapped by bosonization to a free scalar bosonic field theory, described  in terms of the canonically conjugate fields $\hat \Pi$ and $\hat \phi$ by the integrable Hamiltonian~\cite{calzetta_book}
\begin{equation}
    \hat H_{\rm B} =\int\! {\rm d}x\, \left[ \frac{1}{2}{\hat \Pi}^2+\frac{1}{2}(\partial_x \hat \phi)^2+\frac{1}{2} \frac{e^2}{\pi} \hat \phi^2  \right].
    \label{eq:HBS}
\end{equation}
Within this bosonized description, the field $\hat \phi(x,t)$ represents the electric field, and all its Fourier modes $\tilde{ \phi}(k) $ correspond to decoupled harmonic oscillators. The evolution starting from a false vacuum with a uniform  string of non-vanishing electric field  $\langle \hat \phi(x,t=0)\rangle =\text{const} \neq 0$ represents an excitation of the single uniform mode with $k=0$, and hence the electric field will show uniform periodic 
string inversions around zero, with a frequency $\omega_0=e/\sqrt{\pi}$, where $e$ is the charge of the fermion. 
A non-vanishing value of $m$ leads to the additional potential term  
$-cm\omega_0\cos (2\sqrt{\pi}\hat \phi-\theta)$ in the integrand in Eq.~\eqref{eq:HBS},
such that the resulting total potential shows 
a transition from a shape with a single minimum for $m<m_c$ to two 
symmetric minima for $m>m_c$, analogous to the spontaneous breaking of chiral symmetry on the lattice (see the supplementary information for details). This weak local non-linearity introduced by a small $m$ couples the various Fourier modes and hence induces a weak integrability breaking. 
In this case, the uniform string inversions of the electric field evolving from a false vacuum configuration with $\langle \tilde{\phi}(k=0) \rangle \neq 0$ are expected to be superseded by slow thermalization processes at long times (see, e.g., Ref. \onlinecite{ParisiMetastability}).

We suggest that a remnant of this slow dynamics induced by the underlying integrable field theory may persist in lattice versions of this gauge theory as long as initial states with a well-defined continuum limit are considered. With the latter, we mean states whose field configuration is smooth at the level of the lattice spacing: for our case here, the two Neel states represent the smoother ones, as they correspond to the bare vacuum of the fermionic fields, and no electric field excitations. 
At a qualitative level, the effect of integrability-breaking induced by lattice effects is expected to be much weaker in the small Hilbert space sector involving uniform excitations with $k=0$ only, where the long-lived string inversion dynamics take place. The number of states in this sector grows linearly with the lattice size $L$ and their energy spans an extensive range, in agreement with the characteristics of ``many-body quantum scars'', see Ref.~\onlinecite{Papic2018short} and Sec.~\ref{sec:scars} below.

\begin{figure}[t]
\includegraphics[width=0.48\textwidth]{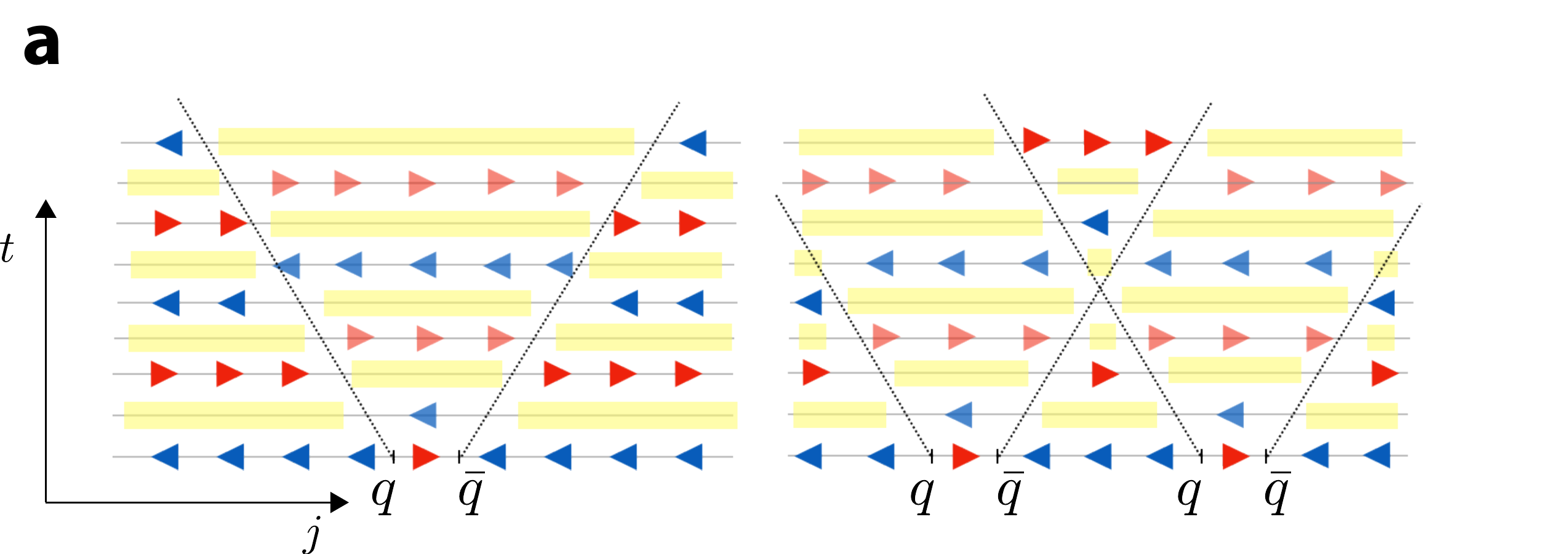}
\includegraphics[width=0.48\textwidth]{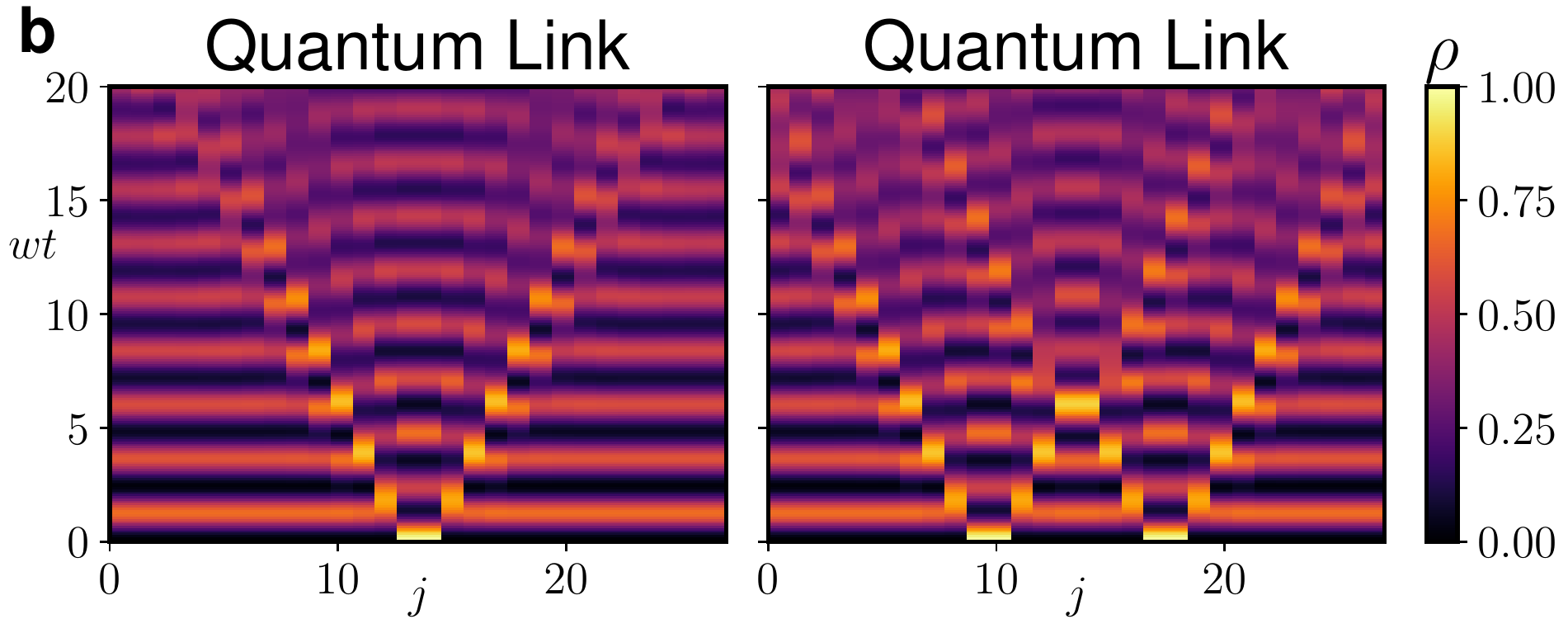}
\includegraphics[width=0.48\textwidth]{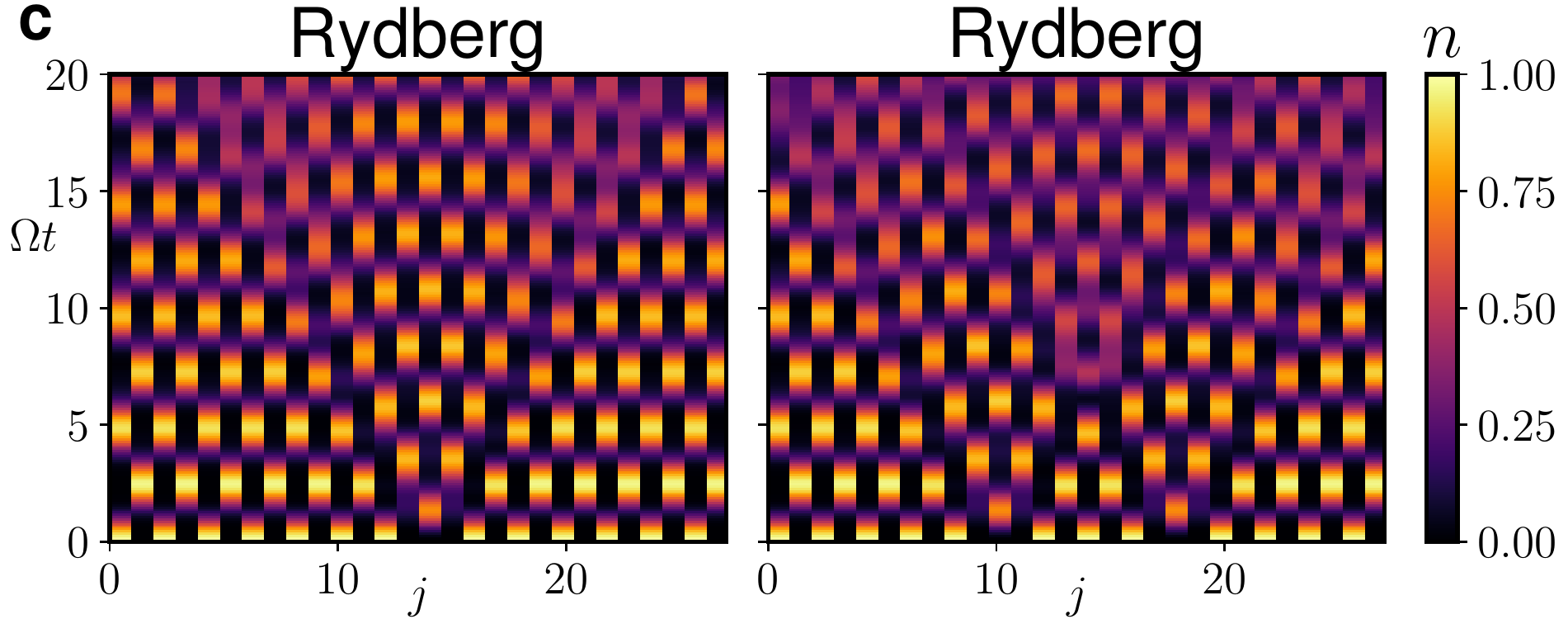}
\caption{%
Slow dynamics of particle-antiparticle pairs.
{\bf a}:
Cartoon states representing the propagation of particle-antiparticle pairs $q$-$\bar q$. The notation is the same as in Fig.~\ref{fig:cartoon}{\bf c}, while the yellow stripes denote regions of space with largest particle density and therefore $\langle \hat E_{j,j+1}\rangle \simeq 0$.
%
%
{\bf b}: 
Evolution of the  particle 
density in the QLM starting from a bare vacuum or "string" state, see Fig.~\ref{fig:cartoon}\textbf{c}, with initial particle-antiparticle pairs.
{\bf c}:
Same as in panel \textbf{b}, but in the Rydberg excitation density representation.
Left column: the oscillations observed in the light-cone shaped region originating from the particles is expected to be out of phase with respect to those 
of the bare vacuum. 
Right column: In the presence of two $q$-$\bar q$ pairs,
an additional change of periodicity is expected in correspondence of elastic scattering. 
%
%
In these simulations, $m=\delta=0$.
}
\label{fig:defects}
\end{figure}

\subsection{Propagation of 
particle-antiparticle pairs}
%
States of the QLM 
corresponding to particle-antiparticle pairs in the bare vacuum can be constructed in Rydberg-atom quantum simulators by preparing two or more defects in a charge-density wave configuration, each corresponding to pairs of adjacent non-excited Rydberg atoms.
 
As an illustration, we discuss how the time-evolution of one or two particle-antiparticle pairs for $m<m_c$ features the emergence of slow dynamics. In Fig.~\ref{fig:defects},  we show the time evolution of both the particle density in the QLM and the corresponding density of excitations in the Rydberg chain, fixing for simplicity $m=0$. The pairs in the initial state break and ballistic spreading 
of quark and antiquark takes place. 
The string inversion dynamics induced by this propagation shows coherent interference patterns with long-lived oscillations. 
Due to retardation effects induced by the constrained dynamics, these oscillations are shifted by half a period with respect to the vacuum oscillation, as captured by second-order perturbation theory.

\begin{center}
\begin{figure}[h]
\includegraphics[width=0.48\textwidth]{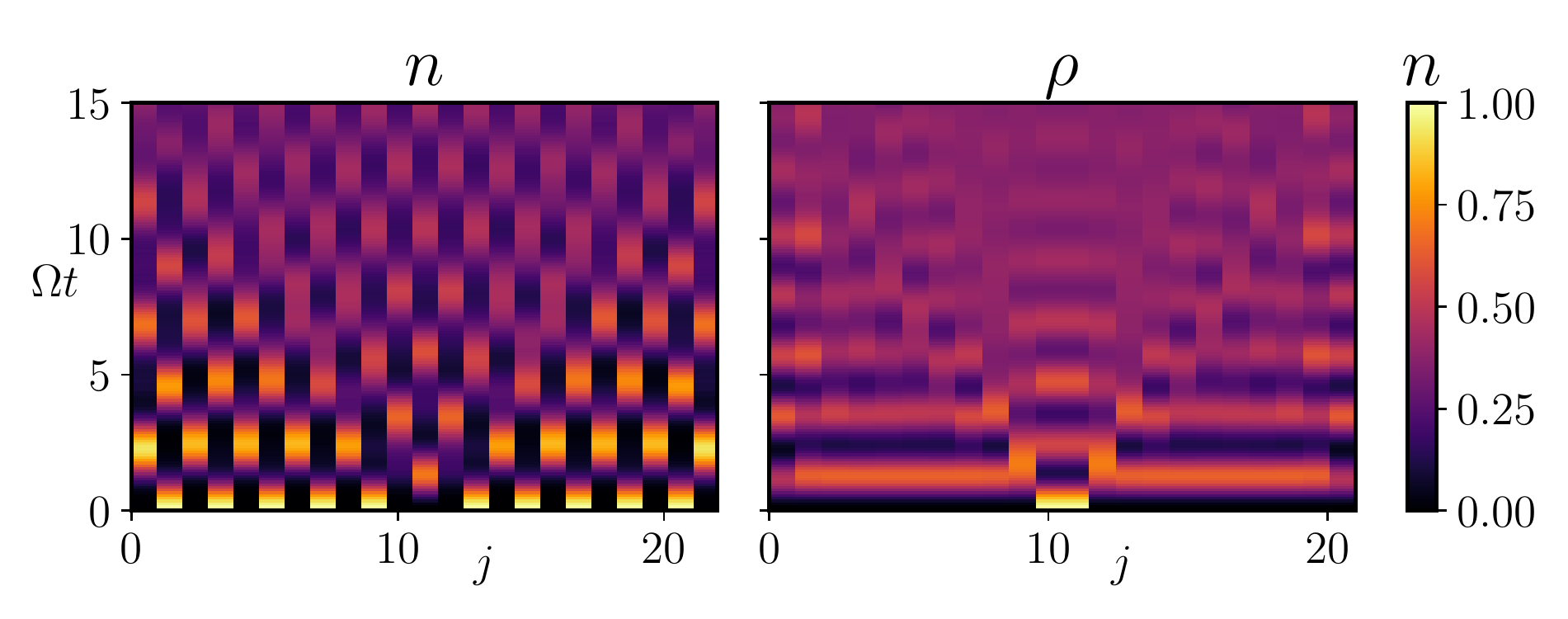}
\caption{Propagation of a particle-antiparticle pair $q$-$\bar q$ with realistic Rydberg interactions. Left panel: density of Rydberg excitations. Right panel: density of particles/antiparticles ($\rho$ in the QLM language). Results are obtained for a chain of $L=23$ sites governed by the realistic Hamiltonian \eqref{eq:Ryd} with  $V_{ij}= V_1 r_{ij}^{-6}$ and no constraints in the Hilbert space. Parameters: $\delta=0$, $V_1/\Omega=25.6$. We checked explicitly that the violation of Rydberg blockade is always small, $\langle n_j n_{j+1} \rangle <  10^{-2}$.}
\label{propagation}
\end{figure}
\end{center}

These unusual dynamics turn out to be robust under experimentally realistic conditions: In Fig.~\ref{propagation} we consider the evolution of a particle-antiparticle pair, the simulated dynamics of which is not constrained to the subspace satisfying $\hat n_j \hat n_{j+1}=0$ and includes the effect of the long-range Rydberg interactions between atoms. The evolution is performed via Krylov subspace techniques in the unconstrained Hilbert space with the Hamiltonian in Eq.~\eqref{eq:Ryd},
with $\delta=0$ and $V_{j,k}=V_1 |j-k|^{-6}$. The value of $V_1/\Omega=25.6$ is the same as considered in Ref.~\onlinecite{Bernien2017}.
The dynamics displayed Fig.~\ref{propagation} is similar to the constrained one in Fig.~\ref{fig:defects}\textbf{b},\textbf{c} at short times, 
after which the effects of having realistic interactions gradually kick in.


\subsection{Spectral properties and bands of non-thermal states}

\label{sec:scars}

 We characterize the anomalous ballistic spreading of particle-antiparticle pairs 
discussed in the previous Section 
in terms of the emergence of corresponding anomalous spectral properties of the FSS model, which generalize those 
recently observed 
\cite{Papic2018short} 
in the special case 
$m=0$, involving families of special energy eigenstates referred to as ``many-body quantum scars''. The latter are constituted by towers of regularly-spaced states in the many-body spectrum with alternating pseudo-momentum $k=0$ and $k=\pi$, characterized by non-thermal expectation values of local observables as well as by anomalously large overlaps with the charge-density wave initial states. The long-lived coherent oscillating behavior has been attributed in Ref.~\onlinecite{Papic2018short} to the existence of these ``scarred'' eigenstates.

\begin{figure}[t]
\includegraphics[width=0.49\textwidth]{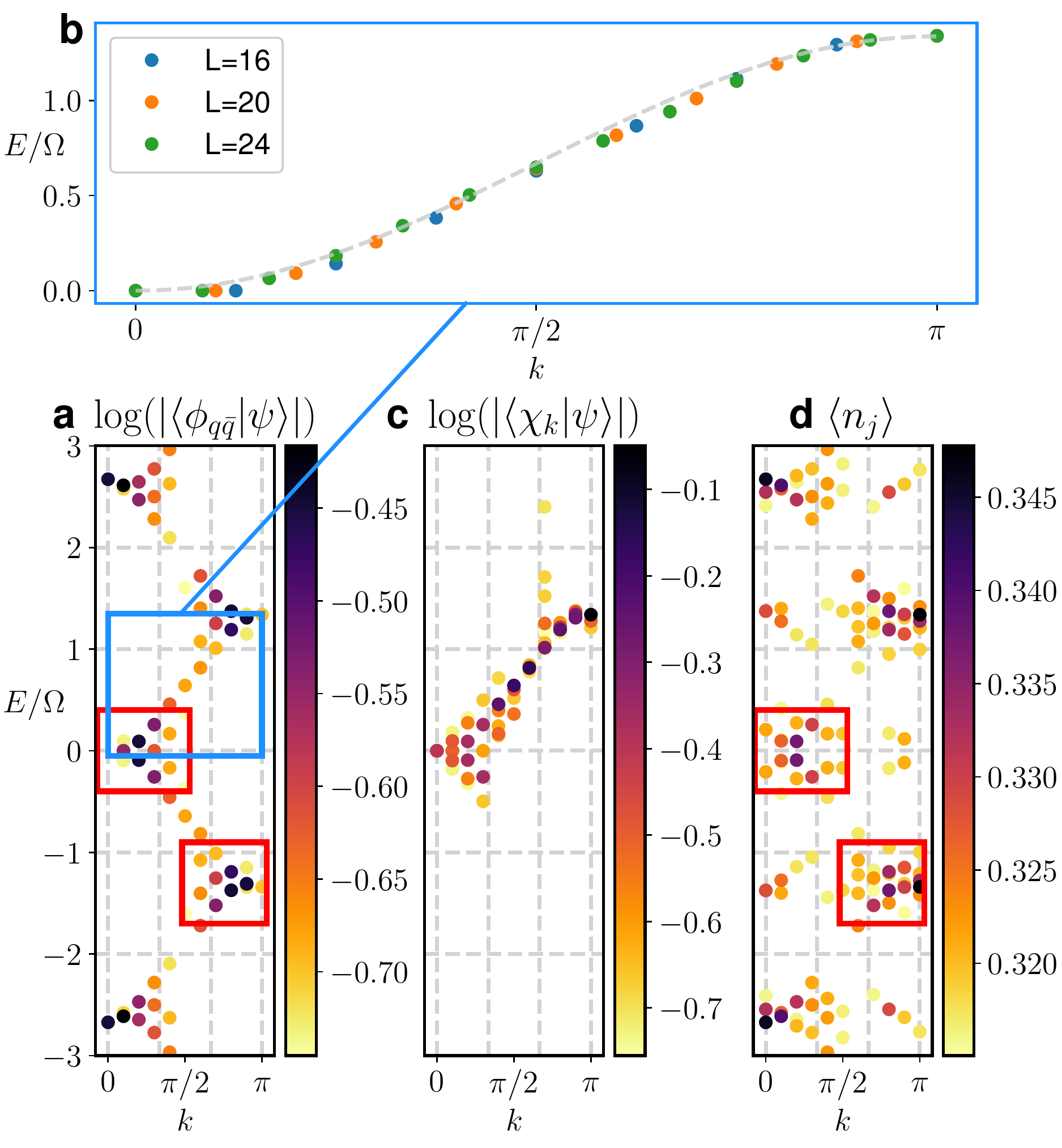}
\caption{%
Emergent quasi-particle description of highly-excited states.
{\bf a}:
Largest overlaps of the initial state   $\ket{\phi_{q\bar q}}$ with a localized defect in a charge-density wave configuration of the Rydberg-atom chain with the energy eigenstates $\ket{\psi}$ of the FSS Hamiltonian ($\delta=0, L=20$) in 
Eq.~\eqref{eq:fss}, as a function of their corresponding momentum and energy. Within the gauge-theory description, the initial state corresponds to having a localized particle-antiparticle pair $q$-$\bar q$.
{\bf b}: 
The eigenstates with the largest overlaps display a regular functional dependence of energy on momentum that is remarkably close to 
a simple cosine band.
{\bf c}:
The largest overlaps of the optimal matrix-product state quasi-particle ansatz $\ket{\chi_k}$ built on an exact eigenstate with zero energy (see the main text) accurately reproduce the corresponding emergent quasi-particle band of panel {\bf a}.
{\bf d}:
Anomalous (non-thermal) expectation values of a local observable in energy eigenstates.
The red boxes highlight the correspondence between the most relevant eigenstates building up $\ket{\phi_{q\bar q}}$ (panel {\bf a}) and the most non-thermal eigenstates (panel {\bf d}).
The emergent spectral structure illustrated in this picture underlies the clean ballistic spreading of particle-antiparticle pairs displayed in 
Fig.~\ref{fig:defects}.
}
\label{fig:spectrum}
\end{figure}

 Fig.~\ref{fig:spectrum}{\bf a} shows that the modulus of the overlap between the energy eigenstate $|\psi\rangle$ with energy $E$ and the above described inhomogeneous states $|\phi_{q\bar{q}}\rangle$ with momentum $k$ clearly identifies a number of special bands of highly-excited 
 energy eigenstates characterized each by an emerging functional relationship $E(k)$. 
 As shown in Fig. \ref{fig:spectrum}{\bf d} some of the states in these bands strongly deviate from the thermal value $\braket{n_j}_{th}\simeq 0.276$. This fact has already been observed in the previously studied quantum-scarred eigenstates, which coincide with the extremal points of these bands at momenta $k=0$ and $k=\pi$.
 A closer inspection of these energy-momentum relations, presented in Fig.~\ref{fig:spectrum}{\bf b}, shows that they are close to cosine-shaped bands, suggesting the emergence of  single-particle excitations in the middle of the many-body energy spectrum.
 
 We further 
 characterize this 
 spectral structure by constructing 
 a quasi-particle variational ansatz $\ket{\chi_k}$ on top of the exact matrix-product-state zero-energy eigenstate of the Hamiltonian \eqref{eq:fss} with $\delta=0$, recently put forward in Ref.~\onlinecite{Motrunich2018} (see the supplementary information).
 As shown in Fig.~\ref{fig:spectrum}{\bf c}, the optimal quasi-particle ansatz has the largest overlap with the states on the energy-momentum bands of special eigenstates closest to zero energy,
 thus reinforcing the above emergent quasi-particle picture.

%

\subsection{Tuning the topological $\theta$-angle in Rydberg experiments}

So far, our discussion has focused exclusively on the relation between Rydberg experiments and the Schwinger model with topological angle $\theta=\pi$. A natural question to ask is whether,  within the present setting, it is possible to realize genuinely confining theories, i.e., generic values of the topological angle $\theta\neq\pi$.

This is possible within the strong coupling limit upon introducing a linear term in the electric field. 
To see this, with reference to the lattice Schwinger model introduced in Sec.~\ref{subs:latticeSchwinger} and notations therein, let us introduce the parameter $\epsilon = \theta/\pi-1$ which  quantifies the deviation of the topological angle from $\pi$. For $|\epsilon|\ll 1$, the two lowest energy states of the electric field have $L_{j,j+1}=0,+1$ on each bond. In order to keep the structure of the Hilbert space compatible with the FSS model, one requires the energy gap $\Delta=2J(1+\epsilon)$ of the next excited state (either $L_{j,j+1}=2$ or $L_{j,j+1}=-1$ depending on the sign of $\epsilon$) 
to be much larger than that separating the first two, i.e., $\sigma=J\epsilon$. 
Accordingly, the lattice Schwinger model with strong $J\gg \Omega, m$ and with a topological angle $\theta = \pi(1 + \sigma/J) $ is efficiently approximated by the QLM with an additional term linear in the electric field and proportional to $\sigma$. The confining nature of the potential can be intuitively understood as follows: starting form the bare vacuum (the ``string'' state in Fig.~\ref{fig:cartoon}), creating and separating a particle-antiparticle pair at a distance $\ell$ entails 
the creation, between the two, of a string of length $\ell$ with opposite electric field.
The corresponding energy cost 
is proportional to $\ell\sigma$, signalling the confining nature of the potential.

In turn, within the exact mapping outlined in Sec.~\ref{sec:mapping} and illustrated in Fig.~\ref{fig:cartoon}, this $\theta-$angle term corresponds to an additional staggered field in the FSS model, leading to the Hamiltonian:
 \beq
 \label{eq:Rydtheta}
 \hat H_{\text{Ryd}} = \sum_{j=1}^{L}(\Omega \,\hat\sigma^x_j +\delta\, \hat\sigma^z_j) + \sum_{j=1}^{L} (-1)^j \frac{\sigma}{2}\sigma^z_j.
 \eeq
 The new term can be experimentally realized, e.g., by utilizing a position dependent AC Stark shift or, alternatively, a space-dependent detuning on the transition between ground and Rydberg states.

\begin{center}
\begin{figure}[h]
\includegraphics[scale=0.5]{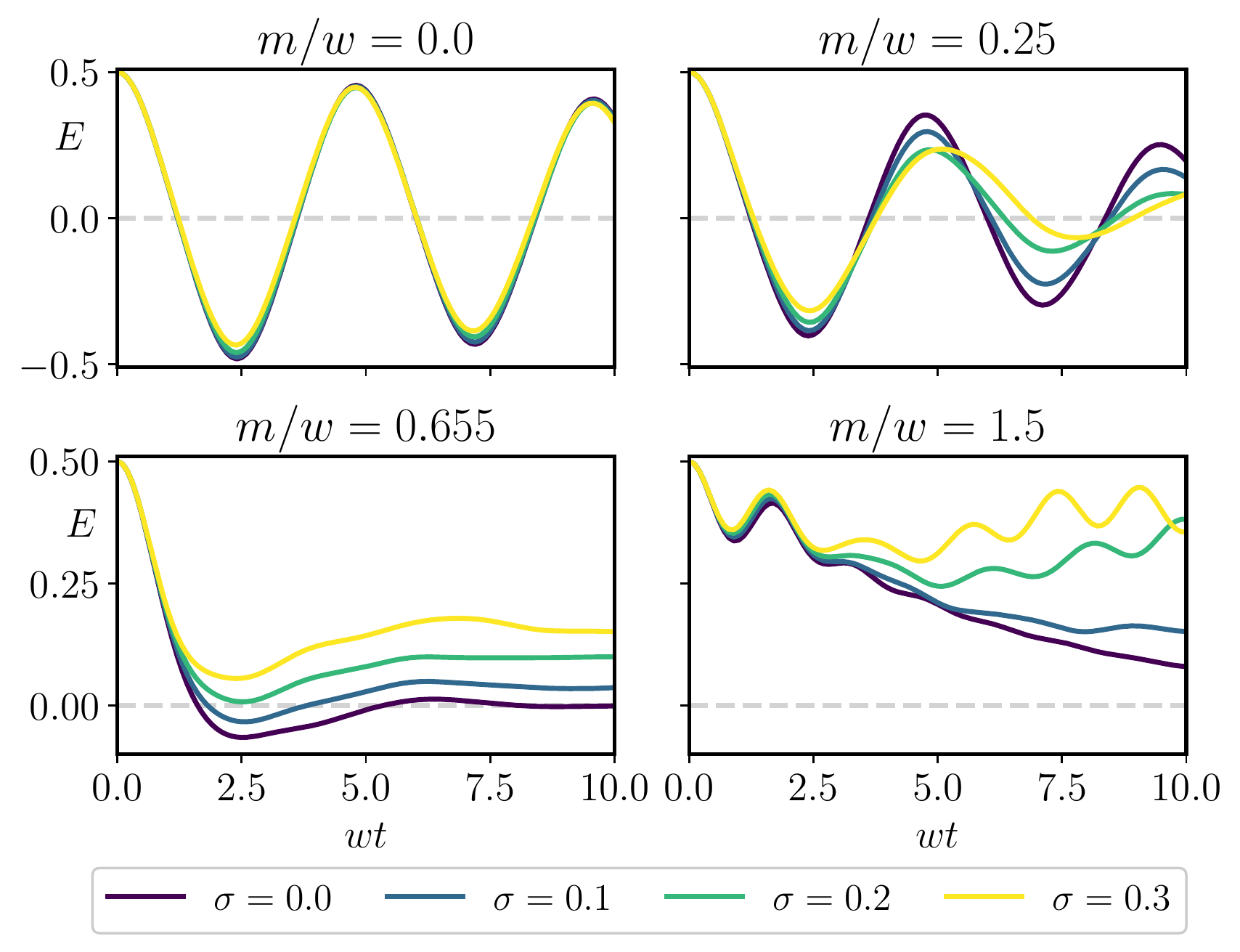}
\caption{Effect of the $\theta-$angle on the dynamics of the electric field from uniform string states of the QLM. Data are shown for a chain of $L=28$ sites, for increasing values of the particle mass $m/w$ and of  the parameter $\sigma$, quantifying the deviation of the $\theta-$angle from $\pi$ (see the main text). Dynamics for $\sigma=0$ correspond to the second column of Fig.~\ref{fig:fss}.
}
\label{fig:thetaangle}
\end{figure}
\end{center}

\begin{center}
\begin{figure}[h]
\includegraphics[width=0.48\textwidth]{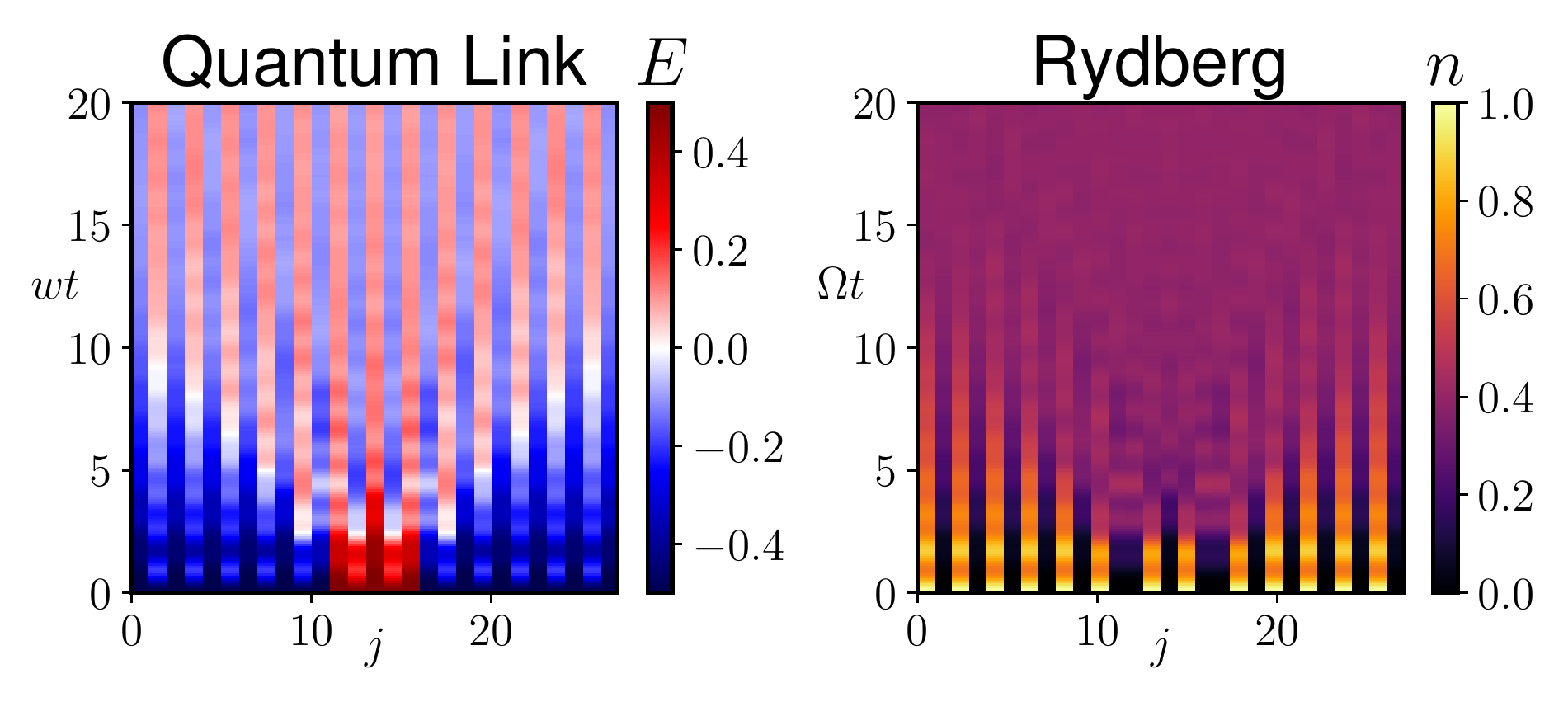} \\
\includegraphics[width=0.48\textwidth]{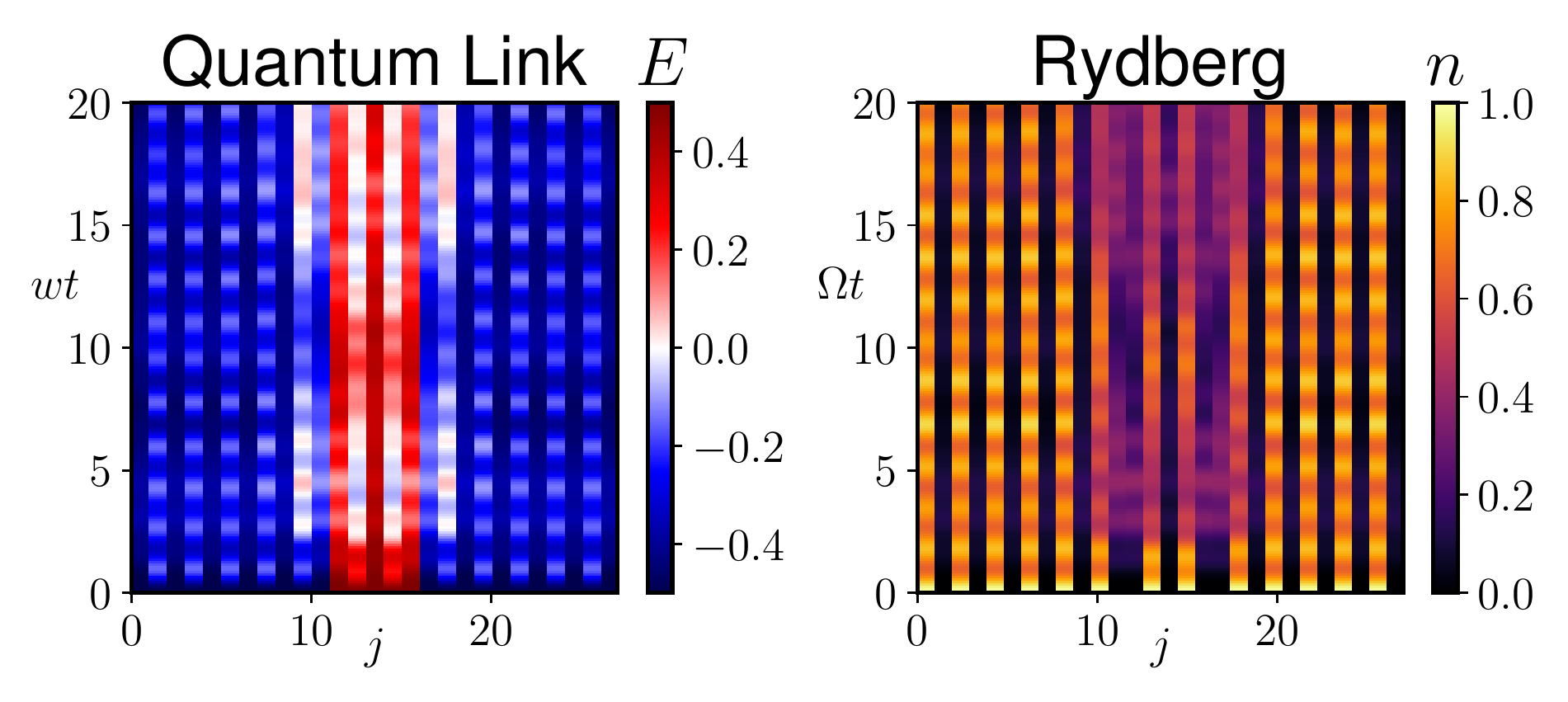} 
\caption{%
$\theta-$angle and string-breaking dynamics. Evolution of a bare particle-antiparticle pair state is displayed in terms of space- and time-dependent electric field in the QLM (left panels) and of the density of excited atoms in the Rydberg array (right panels), with $m=-\delta=1.5\Omega$ and $L=28$. Simulations in the top row have $\sigma=0$, corresponding to the deconfined field theory with $\theta=\pi$. Effects of confinement emerge in the second row, where a non-vanishing 
$\sigma=0.3 \Omega$ stabilizes the electric string.
}
\label{fig:stringbreaking}
\end{figure}
\end{center}

In Fig.~\ref{fig:thetaangle}, we show the effect of the $\theta-$angle on the evolution of the total electric field in the QLM starting from a uniform string state.
As data clearly show, while in the symmetric phase with $m<m_c$, the explicit symmetry breaking caused by the electric field energy imbalance leads to damping of the string inversions, in the broken-symmetry (chiral) phase with $m>m_c$ the effect of confinement is dramatic, causing the persistence of the initial electric string, with small long-lived oscillations.
Focusing on the latter phase, in Fig.~\ref{fig:stringbreaking} we show the dynamical evolution of a finite electric string generated by a particle-antiparticle pair (left panels), at the deconfined point $\theta=\pi$ (top) and in the confined phase with $\theta\ne \pi$ (bottom). The right panels show the same evolution as it would appear in terms of measurements of Rydberg atom excitations. While for $\sigma=0$ nothing prevents the initially localized bare particles to propagate along the chain (top panels), the presence of a linear confining potential proportional to $\sigma$ between them stabilizes the electric string, leading to effective Bloch oscillations of the edges and to a surprisingly long lifetime~\cite{PhysRevB.99.180302} (bottom panels).
This effect signals that confinement can dramatically affect the non-equilibrium dynamics, potentially slowing it down as observed in both gauge theories~\cite{Brenes:2018aa} and statistical mechanics models~\cite{Kormos:2017aa,PhysRevB.99.180302, 2018arXiv180409990J}.

\section{Conclusions}
We proved that the large-scale quantum simulation of lattice gauge theories has already been achieved in state-of-the-art experiments with Rydberg atoms, as it can be realized by establishing a mapping between a $U(1)$ gauge theory and Rydberg atom arrays. At the theoretical level, we showed that this novel interpretation provides additional insights into the exotic dynamics observed in experiments, linking it to archetypal phenomena in particle physics. This immediately implies
their generality and applicability to a wide variety of model Hamiltonians within experimental reach. 
We expect that future studies can further deepen the connection between the statistical mechanics description of such behaviour and its gauge-theoretic interpretation, for instance, elucidating the effects of non-thermal states~\cite{Papic2018short, Papic2018long, Motrunich2018, Pichler2018TDVP} and emergent integrability~\cite{Choi2018SU2, Khemani2018}, and the role of confinement in slowing down the dynamics~\cite{Kormos:2017aa,Brenes:2018aa,PhysRevB.99.180302,2018arXiv180409990J,GlassyDynamicsQuark}. At the experimental level, our findings immediately motivate further experiments along this direction, that can probe different aspects of gauge theories, such as the decay of unstable particle-antiparticle states after a quench, and might be combined with other quantum information protocols~\cite{2018arXiv181003421K}.
The strategy we propose here is based on the elimination of the matter degrees of freedom by exploiting Gauss law: This method does not rely on the specific formulation of the model and is in principle applicable to other lattice gauge theories, including theories with non-Abelian gauge symmetries and in higher dimensions. For a recent work along these lines in the context of non-Abelian theories with finite-dimensional link-Hilbert spaces, see Ref.~\onlinecite{Zohar:aa}.
After the present analysis, the experiments performed in Ref.~\onlinecite{Bernien2017} represent a step-stone toward the ambitious realization of non-Abelian gauge theories in three spatial dimension, which remains an outstanding quest~\cite{Wiese:2013kk,Preskill:2018aa}.

\acknowledgments
We thank M. Aidelsburger, J. Berges, P. Calabrese, M. Collura, A. Dabholkar, P. Hauke, R. Konik, Z. Papic, A. Rudra, and A. Scardicchio for fruitful discussions. MD thanks D. Banerjee, S. Montangero, E. Rico, U.-J. Wiese, and P. Zoller for collaboration on related works, and M. Lukin and H. Pichler for insightful discussions and correspondence. This work is partly supported by the ERC under grant number 758329 (AGEnTh), and has received funding from the European Union's Horizon 2020 research and innovation programme under grant agreement No 817482.

\appendix

\section{Entanglement evolution in the FSS model}
We consider the FSS model defined in Eq.~(2) of the main text and we investigate the time evolution of the bipartite entanglement entropy $S(t)$ of the chain. We consider as initial state the CDW, which is equivalent to considering the QLM evolving from one of the two uniform string configurations, see Fig.~1
in the main text.
In order to determine $S$, we compute the time-dependent reduced density matrix $\hat\rho_\textrm{R}(t)$ of a subsystem consisting of $L/2$ consecutive sites of the chain, by tracing out the degrees of freedom of the remaining complementary $L/2$ sites. In these terms, the von Neumann entanglement entropy is defined by
    $S(t)=-\mathrm{Tr} [\hat\rho_\textrm{R}(t) \ln \hat\rho_\textrm{R}(t)]$.

\begin{figure}[h]
\includegraphics[scale=0.44]{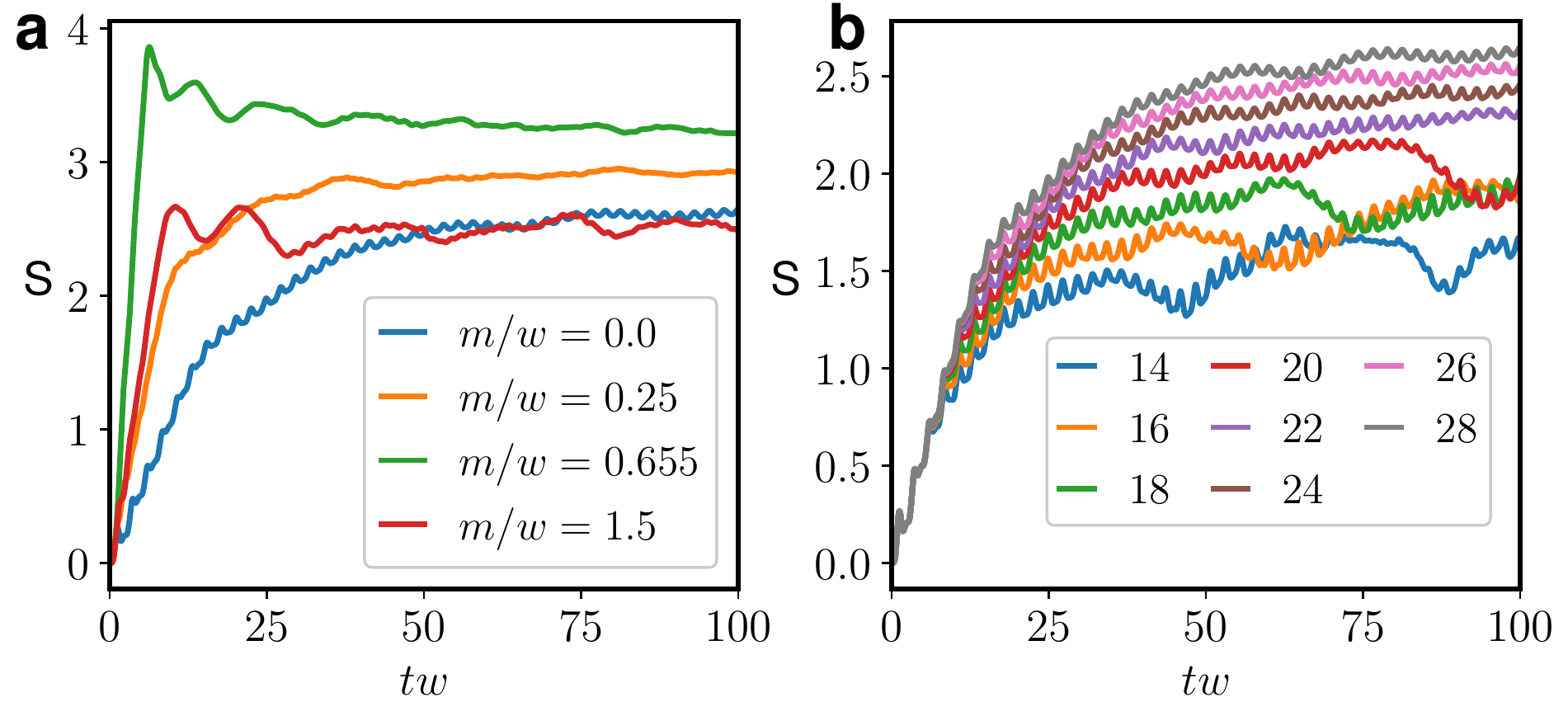} 
\caption{%
Growth of entanglement entropy in the FSS model.
{\bf a:} Growth of the half-chain entanglement entropy for different values of the particle mass $m$. Initial state is CDW/string, and $L=28$. 
{\bf b:} Growth of entanglement entropy for different sizes $L$. Initial state is CDW/string, and $m=0$.
%
%
}
\label{fig:entropy}
\end{figure}

Figures~\ref{fig:entropy}{\bf a} and \ref{fig:entropy}{\bf b} show the evolution of $S$ for various values of the mass $m$ 
and of the chain length $L$, respectively. 
Information spreading is directly tied to particle production: it is fast at the critical point $m=m_c$ (green curve in Fig.~\ref{fig:entropy}{\bf a}, with $m_c/w = 0.655$, see the main text) or above it $m>m_c$ (red curve), where particles are not confined. For $m<m_c$ (yellow and blue curves), instead, it slows down considerably, as was already observed in the spin-$1$ QLM~\cite{Pichler:2016it}. For $m/\omega=0$ the change in the original slope of the curve which occurs around $t\omega\simeq 12$ is due to a 
finite-volume effect, as demonstrated in Fig.~\ref{fig:entropy}{\bf b}, where such a change progressively disappears upon increasing $L$. In all cases, the fast oscillations correspond to different stages of pair production.
%

\section{Spectral properties of the FSS model}

\begin{figure}[b]
\includegraphics[width=0.48\textwidth]{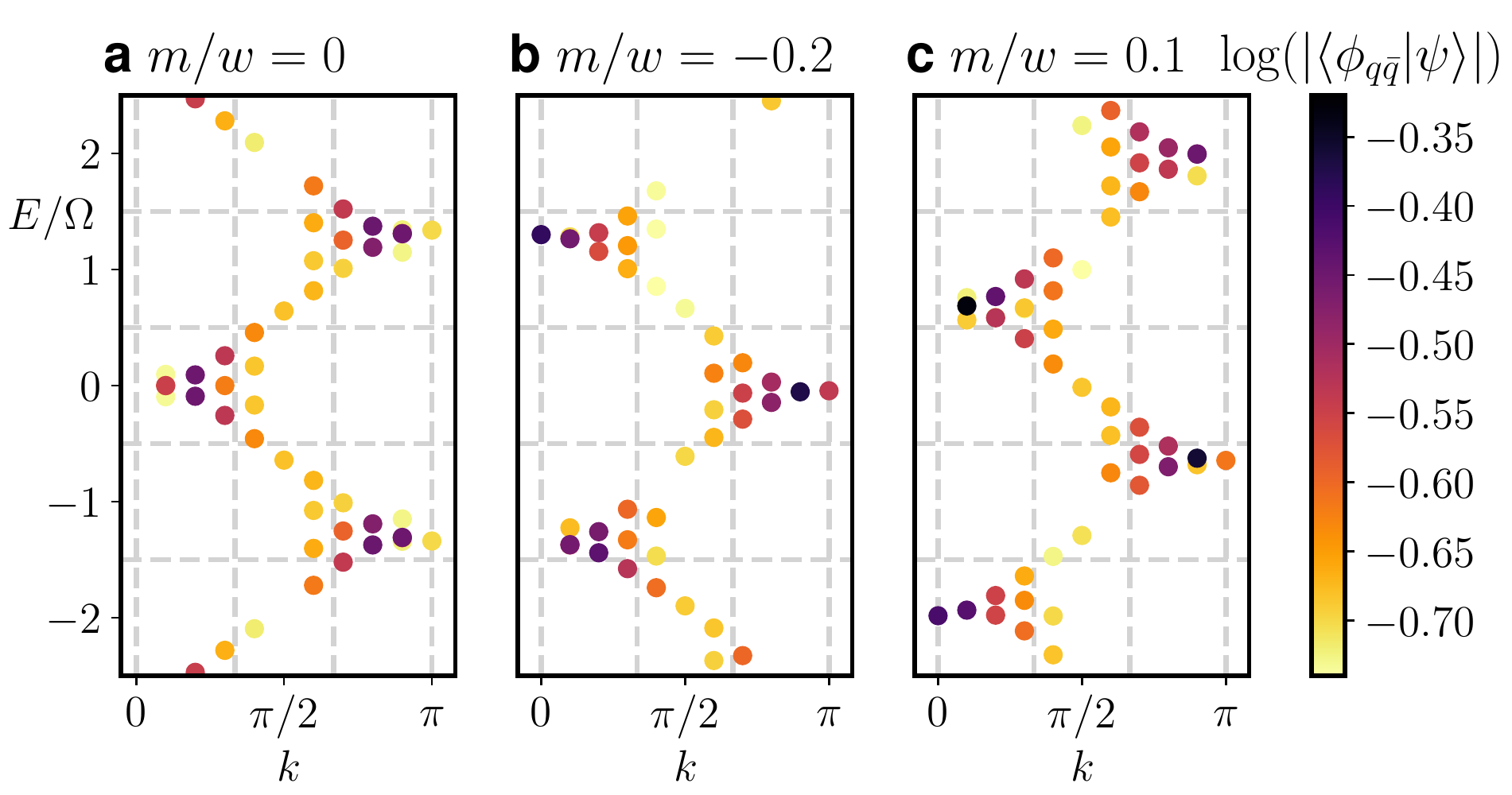}
\caption{%
Robustness of the spectral structure.
Energy-momentum relation of eigenstates around $E=0$ for $L=20$. For each eigenstate $\ket{\psi}$, the colour indicates the value of $\log_{10}{\left(|\braket{\psi|\phi_{q\bar{q}}}|\right)}$ (eigenstates with smallest overlaps are not plotted). The dispersion observed for $m/w=0$ (panel {\bf a}) is shifted but persists when we introduce a non zero mass (panels {\bf b} and {\bf c}).
}
\label{fig:sm1}
\end{figure}

\textit{Robustness of the spectral structure} ---
As shown in the main text, the FSS model for $m=0$ features the emergence of regular structures in the middle of the spectrum {in terms of energy-momentum bands}. 
We here show that these structures are generically present for sufficiently small values of $|m/w|$.
 Figure~\ref{fig:sm1} shows the energy-momentum relation of the eigenstates which have largest overlaps with the inhomogeneous state $|\phi_{q\bar{q}}\rangle$ defined in the main text. For $m/w=0.1$ and $m/w=-0.2$, {similar dispersion relations to the case $m/w=0$ are observed, 
 the main difference being an overall energy shift}.

 \textit{Quasi-particle ansatz for emergent excitations} --- {In order to obtain physical intuition on the emergence of regular energy-momentum bands in highly-excited states which govern the non-equilibrium evolution of 
 localized defects}, we propose the following wavefunction
 \beq
 \ket{\chi_k}=\sum_{j=1}^L e^{-ikj} \hat O_{j-1,j,j+1} \ket{\Phi_{k=0}},
 \label{eq:ansatz}
 \eeq
 where $\ket{\Phi_{k=0}}$ is the exact eigenstate found in Ref.~\cite{Motrunich2018} with momentum $k=0$ and energy $0$, and $\hat O_{j-1,j,j+1}$ is a three-site operator depending on a number of variational parameters. Due to the constraints, the space where this operator acts is reduced from dimension $2^3$ to $5$. 
 The
 inversion symmetry with respect to site $j$
 reduces the number or free variational parameters in $\hat O_{j-1,j,j+1}$ to $11$. We choose a basis of operators $\{\hat M^\alpha_{j-1,j,j+1}\}_{\alpha=1}^{11}$ for parameterizing $\hat O_{j-1,j,j+1}$ and define
 \begin{equation}
     \ket{\phi_k^\alpha}=\sum_{j=1}^L e^{-ikj} \hat M^\alpha_{j-1,j,j+1} \ket{\Phi_{k=0}}.
 \end{equation}

 For each $k$, we minimize the energy variance in the space spanned by the states $\ket{\phi_k^\alpha}$. To this aim, we compute the three matrices  $N^k_{\alpha\beta}=\braket{\phi_k^\alpha|\phi_k^\beta}$,
 $P^k_{\alpha\beta}=\braket{\phi_k^\alpha|\hat H| \phi_k^\beta}$,
 $Q^k_{\alpha\beta}=\braket{\phi_k^\alpha|\hat H^2|\phi_k^\beta}$.
   In order to prevent numerical issues in the minimization, we diagonalize the matrix of the norms $N_k$ and we compute the (rectangular) matrix $U^k$ whose columns are the eigenvectors of $N^k$ having non-zero eigenvalues. We then find the vector $\bm{c_k}=\left(c_k^1, \dots, c_k^{m}\right)$ that minimizes
 \begin{equation}
     \sigma^2_{\hat H} =\frac{\bm{c_k}^\dagger {U^k}^\dagger Q^k U^k \bm{c_k}}{\bm{c_k}^\dagger {U^k}^\dagger N^k  {U^k}\bm{c_k}}- \left(\frac{\bm{c_k}^\dagger  {U^k}^\dagger P^k  {U^k} \bm{c_k}}{\bm{c_k}^\dagger  {U^k}^\dagger N^k  {U^k} \bm{c_k}}\right)^2.
 \end{equation}
 Note that by introducing the matrix $U^k$ we restrict the minimization to states with non-zero norms, thus further reducing the number of variational parameters to $m(k)\le 11$. The optimal wavefunction is then obtained as
 \begin{equation}
     \ket{\chi_k}=\sum_{\alpha=1}^{11}\sum_{\beta=1}^m U^k_{\alpha\beta}c_k^\beta \ket{\phi_k^\beta}.
 \end{equation}

\section{Dynamics of the Schwinger model}

\textit{Mapping onto a long-range interacting spin chain} --- The lattice Schwinger model in Eq.~(3) of the main text\textbf{} in the gauge-invariant subspace spanned by wavefunctions $\ket{\psi}$ which satisfy the Gauss laws $\hat G_j\ket{\psi}=0$, can be conveniently simulated by exactly mapping it onto an unconstrained chain of spin-$1/2$ degrees of freedom in the case of open boundary conditions~\cite{Banks1976}. These spins are obtained from the fermionic operators via a combination of a Jordan-Wigner transformation and a gauge transformation, 
expressed as
\beq
\hat \Phi_j = \prod_{l=1}^{j-1}\left(\hat\sigma_l^z \hat U_{l,l+1}^\dagger\right)\hat \sigma_j^-.
\eeq
This transformation decouples spins and gauge degrees of freedom, and thus
the Hamiltonian~(3) in the main text takes the form
\begin{multline}
   \hat{H}=-w\sum_{j=1}^{L-1}(\hat\sigma_j^+\hat \sigma^-_{j+1} + \textnormal{h.c.})+ \frac{m}{2}\sum_{j=1}^L(-1)^j\hat\sigma^z_j \\+ J\sum_{j=1}^{L-1} \hat E^{\mathstrut 2 }_{j,j+1} 
   .
\label{ham}
\end{multline}
The electric field can be rewritten in terms of the spin operators by means of the Gauss law,
\begin{equation}
\hat E_{j,j+1}=\frac{1}{2}\sum_{l=1}^j\left[\hat \sigma_l^z+(-1)^l\right]-\alpha.
\label{electric}
\end{equation}
Inserting Eq.~\eqref{electric} into Eq.~\eqref{ham} we obtain three additional terms: a long-range spin-spin interaction corresponding to a Coulomb interaction, a local energy offset that modifies the effective mass of the fermions and a linear potential given by the constant background field. The electric field part of the Hamiltonian can be cast in the form:
\begin{multline}\label{eq:HE}
       \hat{H}_{\textnormal{lat}}^E=\frac{J}{2}\sum_{n=1}^{L-2}\sum_{l=n+1}^{L-1}(L-l)\hat{\sigma}^z_n\hat{\sigma}^z_l \\
       -\frac{J}{4}\sum_{n=1}^{L-1}[1-(-1)^n]\sum_{l=1}^n\hat{\sigma}^z_l-J\alpha \sum_{j=1}^{L-1}(L-j)\hat \sigma^z_j.
\end{multline}
In this form, the non-equilibrium dynamics of the lattice Schwinger model can be efficiently simulated with standard algorithms of quantum many-body physics.

The origin of long-range spin-spin interactions as a consequence of the linear confining Coulomb potential in one spatial dimension is made more evident when Eq.~\eqref{eq:HE} is formulated in terms of the charges $\hat Q_j=\left[\hat \sigma_j^z+(-1)^j\right]/2$ \cite{CiracGaussian}. In the neutral charge sector where $\sum_{j=1}^L\hat Q_j=0$ we have

\begin{multline}\label{eq:HEcoulomb}
    \hat H^E_{lat}=
    -J\sum_{j=1}^{L-1}\sum_{k=j+1}^{L}(k-j)\, \hat Q_j\hat Q_k\\
    -J \sum_{j=1}^{L}(L+1-j)\,\alpha\hat Q_j+J \sum_{j=1}^{L}j\,\alpha\hat Q_j.
\end{multline}
The first term describes the
Coulomb interaction between charges, while the remaining two terms can be interpreted as interactions with two static charges $-\alpha$ and $\alpha$, placed at the boundaries of the chain (sites $0$ and $L+1$ respectively) and effectively producing the constant background field.

\textit{Continuum limit of the massive Schwinger model} --- 
The massive Schwinger model {briefly introduced in the main text} describes the quantum electrodynamics of fermions of mass $m$ and charge $e$ in $1+1$ dimensions. Its Lagrangian density is
\begin{equation}\label{eq:lagrangian}
\mathcal{L}=-\frac{1}{4}F_{\mu\nu}F^{\mu\nu}+\bar{\psi}(i\fsl{\partial} -e \fsl{A}-m )\psi
\end{equation}
 where $F_{\mu\nu}=\partial_\mu A_\nu-\partial_\mu A_\nu$. The indices $\mu,\nu=0,1$ indicate respectively the time and space directions, and the slash notation indicates contraction with the Dirac matrices $\gamma_\mu$. This model
 can be formulated in terms of a bosonic field $\phi$~\cite{Schwinger2}. We 
 {briefly recall here the main points of} the derivation of the bosonic Hamiltonian obtained in Ref. \onlinecite{Coleman1976239}.
 
 In the Coulomb gauge ($A_1=0$), the Euler-Lagrange equation for $A_0$ yields
 \begin{equation}\label{eq:continuumGauss}
     \partial^2_1 A_0= -ej_0
 \end{equation}
 where $j_0=\psi^\dagger \psi$ is the charge density. Integrating Eq.~\eqref{eq:continuumGauss}, we obtain the continuum version of Eq.~\eqref{electric},
 \begin{equation}\label{eq:field}
     F_{01}=-\partial_1 A_0=e\partial_1^{-1}j_0 +F
 \end{equation}
 where $F$ is a number, representing a classical background field.
  The Hamiltonian density obtained from the Lagrangian \eqref{eq:lagrangian} has the form
 \begin{equation}\label{eq:Hd}
      \mathcal{H}=\bar{\psi}(i\gamma_1\partial_1+m )\psi+ \frac{1}{2}{F_{01}}^2.
 \end{equation}
 
    The interacting Hamiltonian for the fermions can be formulated using Eq.~\eqref{eq:field} to integrate out the gauge fields. Integrating by parts in the zero charge sector, i.e., $\int dx\; j_0(x)=0$, we obtain
 \begin{multline}
     H =\int dx\; \bar{\psi}(i\gamma_1\partial_1+m )\psi\\
     -\frac{e^2}{4}\int dx\; dy\; j_0(x)j_0(y)|x-y|
     -eF\int dx \; xj_0(x).
 \end{multline}
 Similarly to the lattice version of this model [cf. Eqs.~\eqref{ham} and \eqref{eq:HEcoulomb}], the resulting Hamiltonian contains the energy of massive free fermions, the Coulomb interaction between charges (which increases linearly in one spatial dimension) and the interactions between the charges and the background field. 
 
 The method of bosonization can be applied, by noting that in $1+1$ dimensions the conserved vector field $j^\mu=\bar\psi \gamma^\mu \psi$ can be written as the curl of a scalar field $\phi$
 \begin{equation}
     j_\mu =\pi^{-1/2}\epsilon_{\mu\nu}\partial^\nu \phi.
 \end{equation}
 By substituting in Eq.~\eqref{eq:field} we get
 \begin{equation}\label{eq:F01}
     F_{01}=e\pi^{-1/2}\phi+F,
 \end{equation}
 and, from the results obtained for a free massive Dirac field~\cite{Coleman_sinegordon}, we know
 \begin{multline}\label{eq:sg}
    \bar{\psi}(i\gamma_1\partial_1+m )\psi \rightarrow N_m\left[\frac{1}{2}{ \Pi}^2+\frac{1}{2}(\partial_1 \phi)^2+\right.\\
    \left.-cm^2\cos(2\pi^{1/2}\phi)\right].
\end{multline}
where $c=e^\gamma/(2\pi)$, $\gamma\simeq 0.577$ is the Euler constant and $N_m$ indicates normal ordering with respect to the mass $m$. Inserting Eqs.~\eqref{eq:F01} and \eqref{eq:sg} in Eq.~\eqref{eq:Hd}, the Hamiltonian density reads
\begin{multline}
\mathcal{H}= N_m\left[\frac{1}{2}{ \Pi}^2+\frac{1}{2}(\partial_1 \phi)^2
    -cm^2\cos(2\pi^{1/2}\phi)+\right.\\
    \left.+\frac{e^2}{2\pi}\left(\phi+\frac{\pi^{1/2}F}{e}\right)^2\right].
\end{multline}
By shifting the field $\phi\rightarrow \phi-\pi^{1/2}F/e$ and defining a new normal ordering with respect to the mass $\mu=\pi^{-1/2}e$, we finally obtain
\begin{multline}
\mathcal{H}= N_\mu\left[\frac{1}{2}{ \Pi}^2+\frac{1}{2}(\partial_1 \phi)^2-cm\mu\cos(2\pi^{1/2}\phi-\theta)+
    \right.\\
    \left.+\frac{\mu^2}{2}\phi^2\right]
\end{multline}
where $\theta=2\pi F/e$.
{The latter form connects with the discussion in the main text -- cf. Eq. (5) therein.}

\bibliography{bib}

\end{document}